\documentclass[aip,jcp,preprint,superscriptaddress,nofootinbib]{revtex4-1}
\usepackage{silence}
\usepackage[english]{babel}
\usepackage[T1]{fontenc}
\usepackage[utf8]{inputenc}
\usepackage{microtype}

\usepackage{amsmath}
\usepackage{amssymb}
\usepackage{textcomp}
\usepackage{textgreek}
\usepackage{physics}
\usepackage{upgreek}
\usepackage{mathtools}
\usepackage{bbm}
\usepackage{bm}
\usepackage{mathrsfs}
\usepackage{nicefrac}

\usepackage{graphicx}
\usepackage{xcolor}
\usepackage{float}
\usepackage{booktabs}
\usepackage{enumitem}
\usepackage[colorlinks]{hyperref}
\usepackage{braket}
\usepackage{array}
\usepackage{siunitx}
\usepackage{threeparttable}
\usepackage{pifont}
\usepackage{lipsum}
\usepackage[labelformat=simple]{subcaption}

\newcommand{\bottomstrut}[1]{\rule[#1]{0pt}{0pt}}

\newcommand\Tstrut{\rule{0pt}{2.4ex}}

\newcommand\TTstrut{\rule{0pt}{3.0ex}}

\newcommand{\mrm}[1]{\mathrm{#1}}
\newcommand{\mbf}[1]{\mathbf{#1}}
\newcommand{\nn}{\nonumber\\}
\newcommand{\trans}{\mspace{-3mu}\mathsf{T}}
\newcommand{\subA}{\textsc{a}}
\newcommand{\subV}{\textsc{v}}
\newcommand{\subO}{\textsc{o}}
\newcommand{\subR}{\textsc{r}}
\newcommand{\subS}{\textsc{s}}
\newcommand{\bigbraket}[2]{\Big\langle #1 \mspace{2mu} \Big\vert \mspace{1.5mu} #2 \Big\rangle}

\newcommand{\2}{\mspace{-2mu}}
\newcommand{\elm}[3]{\langle #1 \vert #2\vert #3 \rangle}

\DeclareRobustCommand{\raisedrho}{{\mathpalette\irrho\relax}}
\newcommand{\irrho}[2]{\raisebox{1pt}{$#1\rho$}}

\makeatletter

\def\env@dcases#1{%
  \let\@ifnextchar\new@ifnextchar
  \left\lbrace\def\arraystretch{1.2}%
  \array{@{}#1@{\quad}l@{}}}
\makeatother

\newcommand*\xbar[1]{%
  \hbox{%
  \kern0.3em%
    \vbox{%
      \hrule height 0.5pt 
      \kern0.3ex
      \hbox{%
        \kern-0.3em
        \ensuremath{#1}%
        \kern-0.2em
      }%
    }%
    \kern0.2em%
  }%
} 

\DeclareFontFamily{OT1}{pzc}{}
\DeclareFontShape{OT1}{pzc}{m}{it}{<-> s * [1.10] pzcmi7t}{}
\DeclareMathAlphabet{\mathpzc}{OT1}{pzc}{m}{it}

\newcommand{\Eplain}[3]{E^{#1}_{{#2} {#3}}}
\newcommand{\Etilde}[3]{\tilde{E}^{#1}_{{#2} {#3}}}
\newcommand{\Ebreve}[3]{\breve{E}^{#1}_{{#2} {#3}}}

\newcommand{\Etildeprime}[3]{\tilde{E}^{#1'}_{{#2} {#3}}}

\newcommand{\kappaplain}[3]{\kappa^{#1}_{{#2} {#3}}}
\newcommand{\kappaplaindot}[3]{\dot{\kappa}^{#1}_{{#2} {#3}}}

\newcommand{\sigmaplain}[3]{\sigma^{#1}_{{#2} {#3}}}

\newcommand{\gplain}[3]{g^{#1}_{{#2} {#3}}}

\newcommand{\FFcheckplain}[1]{\check{\mbf{F}}^{#1}}

\newcommand{\FFcheckprimedagger}[1]{\check{\mbf{F}}^{\prime#1\dagger}}
\newcommand{\FFtildeplain}[1]{\tilde{\mbf{F}}^{#1}}

\newcommand{\FFtildeprimedagger}[1]{\tilde{\mbf{F}}^{\prime#1\dagger}}

\newcommand{\Fcheckplain}[3]{\check{F}^{#1}_{{#2} {#3}}}

\newcommand{\Fcheckprime}[3]{\check{F}^{\prime#1}_{{#2} {#3}}}
\newcommand{\Ftildeplain}[3]{\tilde{F}^{#1}_{{#2} {#3}}}
\newcommand{\Fbreveplain}[3]{\breve{F}^{#1}_{{#2} {#3}}}
\newcommand{\Ftildeprime}[3]{\tilde{F}^{\prime#1}_{{#2} {#3}}}
\newcommand{\Fbreveprime}[3]{\breve{F}^{\prime#1}_{{#2} {#3}}}

\newcommand{\fplain}[3]{f_{(#1  {#2} {#3})}}
\newcommand{\fprime}[3]{f'_{(#1  {#2} {#3})}}

\newcommand{\rrho}[3]{\rho^{#1}_{#2 #3}}

\newcommand{\dplain}[5]{d^{#1}_{({#2} {#3}) ({#4} {#5})}}

\newcommand{\Cprime}[6]{C'_{ ({#1} {#2}  {#3}) ({#4'} {#5} {#6}) } }

\newcommand{\Cplain}[6]{C_{ ({#1} {#2}  {#3}) ({#4'} {#5} {#6}) } }

\newcommand{\Aprime}[4]{A_{ {#1} ({#2'} {#3} {#4}) } }

\newcommand{\btilde}[2]{\tilde{b}^{#1}_{\mspace{-1mu}#2}}
\newcommand{\bbreve}[2]{\breve{b}^{#1}_{\mspace{-1mu}#2}}

\newcommand{\crea}[2]{a^{#1\mspace{-0.5mu}\raisebox{0.3ex}{$\scriptstyle\dagger$}}_{\mspace{-0mu}#2}}
\newcommand{\creatilde}[2]{\tilde{a}^{#1\mspace{-0.5mu}\raisebox{0.3ex}{$\scriptstyle\dagger$}}_{\mspace{-0mu}#2}}
\newcommand{\creabreve}[2]{\breve{a}^{#1\mspace{-0.5mu}\raisebox{0.3ex}{$\scriptstyle\dagger$}}_{\mspace{-0mu}#2}}
\newcommand{\anni}[2]{a^{#1}_{\mspace{-0mu}#2}}
\newcommand{\annitilde}[2]{\tilde{a}^{#1}_{\mspace{-0mu}#2}}

\newcommand{\annitildeelec}[1]{\tilde{a}_{#1}}
\newcommand{\creatildeelec}[1]{\tilde{a}^{\dagger}_{#1}}

\newcommand{\chiplain}[2]{\chi^{#1}_{#2}}
\newcommand{\chiconj}[2]{\chi^{#1*}_{#2}}

\newcommand{\phitilde}[2]{\tilde{\varphi}^{#1}_{#2}}

\newcommand{\phitildeprime}[2]{\tilde{\varphi}^{\prime#1}_{#2}}

\newcommand{\vplain}[3]{V^{#1}_{\mspace{-2mu} #2 #3}}
\newcommand{\vplainconj}[3]{V^{#1*}_{\mspace{-2mu} #2 #3}}
\newcommand{\vplaindot}[3]{\dot{V}^{#1}_{\mspace{-2mu} #2 #3}}
 \usepackage{acro}

\DeclareAcronym{mctdh}{
   short = MCTDH ,
   long = multiconfiguration time-dependent Hartree ,
}

\DeclareAcronym{nomctdh}{
   short = NOMCTDH ,
   long = non-orthogonal \ac{mctdh} ,
}

\DeclareAcronym{gmctdh}{
   short = G-MCTDH ,
   long = Gaussian-based \ac{mctdh} ,
}

\DeclareAcronym{mlgmctdh}{
   short = ML-GMCTDH ,
   long = multilayer Gaussian-based \ac{mctdh} ,
}

\DeclareAcronym{mlmctdh}{
   short = ML-MCTDH ,
   long = multilayer \ac{mctdh} ,
}

\DeclareAcronym{mpsmctdh}{
   short = MPS-MCTDH ,
   long = matrix product state \ac{mctdh} ,
}

\DeclareAcronym{vmcg}{
   short = vMCG ,
   long = variational multiconfiguration Gaussian ,
}

\DeclareAcronym{ms}{
   short = MS ,
   long = multiple spawning ,
}

\DeclareAcronym{ccs}{
   short = CCS ,
   long = coupled coherent states ,
}

\DeclareAcronym{mctdhn}{
   short = MCTDH[\textit{n}] ,
   long = systematically truncated multiconfiguration time-dependent Hartree ,
}

\DeclareAcronym{mrmctdhn}{
  short = MR-MCTDH[\textit{n}] ,
  long = multi-reference truncated multiconfiguration time-dependent Hartree ,
}

\DeclareAcronym{tdh}{
   short = TDH ,
   long = time-dependent Hartree ,
}

\DeclareAcronym{dmrg}{
   short = DMRG ,
   long = density matrix renormalization group,
}

\DeclareAcronym{tddmrg}{
   short = TD-DMRG ,
   long = time-dependent density matrix renormalization group,
}

\DeclareAcronym{scf}{
   short = SCF ,
   long = self-consistent field ,
}

\DeclareAcronym{casscf}{
   short = CASSCF ,
   long = complete active space self-consistent field ,
}

\DeclareAcronym{tdcasscf}{
   short = TD-CASSCF ,
   long = time-dependent \acl{casscf} ,
}

\DeclareAcronym{gasscf}{
   short = CASSCF ,
   long = generalized active space self-consistent field ,
}

\DeclareAcronym{tdgasscf}{
   short = TD-GASSCF ,
   long = time-dependent \acl{gasscf} ,
}

\DeclareAcronym{rasscf}{
   short = RASSCF ,
   long = restricted active space self-consistent field ,
}

\DeclareAcronym{tdrasscf}{
   short = TD-RASSCF ,
   long = time-dependent \acl{rasscf} ,
}

\DeclareAcronym{ormas}{
   short = ORMAS ,
   long = occupation-restricted multiple active space ,
}
\DeclareAcronym{tdormas}{
   short = TD-ORMAS ,
   long = time-dependent \acl{ormas} ,
}

\DeclareAcronym{mctdhf}{
   short = MCTDHF ,
   long = multiconfiguration time-dependent Hartree--Fock ,
}

\DeclareAcronym{occ}{
   short = OCC ,
   long = orbital-optimized coupled cluster ,
}

\DeclareAcronym{bcc}{
   short = BCC ,
   long = Brueckner coupled cluster ,
}

\DeclareAcronym{tdocc}{
   short = TDOCC ,
   long = time-dependent \acl{occ} ,
}

\DeclareAcronym{nocc}{
   short = NOCC ,
   long = non-orthogonal orbital-optimized coupled cluster ,
}

\DeclareAcronym{tdnocc}{
   short = TDNOCC ,
   long = time-dependent \acl{nocc} ,
}

\DeclareAcronym{oatdcc}{
   short = OATDCC ,
   long = orbital-adaptive time-dependent coupled cluster ,
}

\DeclareAcronym{fci}{
   short = FCI ,
   long = full configuration interaction ,
}

\DeclareAcronym{cud}{
   short = CUD ,
   long = closed under de-exciation ,
}

\DeclareAcronym{fsmr}{
   short = FSMR ,
   long = full-space matrix representation ,
}

\DeclareAcronym{hh}{
   short = HH ,
   long = H\'enon-Heiles ,
}

\DeclareAcronym{ho}{
   short = HO ,
   long = harmonic oscillator ,
}

\DeclareAcronym{dop853}{
   short = DOP853 ,
   long = Dormand-Prince 8{(5,3)} ,
}

\DeclareAcronym{sm}{
   short = SM ,
   long = supplementary material ,
}

\DeclareAcronym{vscf}{
   short = VSCF ,
   long = vibrational self-consistent field ,
}

\DeclareAcronym{eom}{
   short = EOM ,
   long = equation of motion ,
   short-plural-form = EOMs ,
   long-plural-form = equations of motion ,
}

\DeclareAcronym{tdvp}{
   short = TDVP ,
   long = time-dependent variational principle
}

\DeclareAcronym{tdse}{
   short = TDSE ,
   long = time-dependent Schr{\"o}dinger equation ,
}

\DeclareAcronym{cc}{
   short = CC ,
   long = coupled cluster ,
}

\DeclareAcronym{vcc}{
   short = VCC ,
   long = vibrational coupled cluster ,
}

\DeclareAcronym{tdvcc}{
   short = TDVCC ,
   long = time-dependent vibrational coupled cluster ,
}

\DeclareAcronym{tdvci}{
   short = TDVCI ,
   long = time-dependent vibrational configuration interaction ,
}

\DeclareAcronym{vci}{
   short = VCI ,
   long = vibrational configuration interaction ,
}

\DeclareAcronym{ci}{
   short = CI ,
   long = configuration interaction ,
}

\DeclareAcronym{tdci}{
   short = CI ,
   long = time-dependent \acl{ci} ,
}

\DeclareAcronym{sq}{
   short = SQ ,
   long = second quantization ,
}

\DeclareAcronym{fq}{
   short = FQ ,
   long = first quantization ,
}

\DeclareAcronym{mc}{
   short = MC ,
   long = mode combination ,
}

\DeclareAcronym{mcr}{
   short = MCR ,
   long = mode combination range ,
   long-plural = s ,
}

\DeclareAcronym{pes}{
   short = PES ,
   long = potential energy surface
}

\DeclareAcronym{svd}{
   short = SVD ,
   long = singular value decomposition ,
}
\DeclareAcronym{adga}{
   short = ADGA ,
   long = adaptive density-guided approach ,
}

\DeclareAcronym{rhs}{
   short = RHS ,
   long = right-hand side ,
}

\DeclareAcronym{lhs}{
   short = LHS ,
   long = left-hand side ,
}

\DeclareAcronym{ivr}{
   short = IVR ,
   long = intramolecular vibrational energy redistribution ,
}

\DeclareAcronym{fft}{
   short = FFT ,
   long = fast Fourier transform ,
}

\DeclareAcronym{spf}{
   short = SPF ,
   long = single-particle function ,
}

\DeclareAcronym{lls}{
   short = LLS ,
   long = linear least squares ,
}

\DeclareAcronym{itnamo}{
   short = ItNaMo ,
   long = iterative natural modal ,
}

\DeclareAcronym{hf}{
   short = HF ,
   long = Hartree--Fock ,
}

\DeclareAcronym{mcscf}{
   short = MCSCF ,
   long = multi-configurational self-consistent field ,
}

\DeclareAcronym{sop}{
   short = SOP ,
   long = sum-of-products ,
}

\DeclareAcronym{mpi}{
   short = MPI ,
   long = message passing interface ,
}

\DeclareAcronym{ode}{
   short = ODE ,
   long  = ordinary differential equation ,
   short-plural = s ,
   long-plural = s ,
   short-indefinite = an ,
   long-indefinite = an ,
   tag = abbrev ,
}

\DeclareAcronym{bch}{
   short = BCH ,
   long = Baker--Campbell--Hausdorff ,
}

\DeclareAcronym{sr}{
   short = SR ,
   long = single-reference ,
}
\DeclareAcronym{mr}{
   short = MR ,
   long = multi-reference ,
}

\DeclareAcronym{dof}{
   short = DOF ,
   long = degree of freedom ,
   short-plural-form = DOFs ,
   long-plural-form = degrees of freedom ,
}

\DeclareAcronym{hp}{
   short = HP ,
   long = Hartree product ,
}

\DeclareAcronym{tdbvp}{
   short = TDBVP ,
   long  = time-dependent bivariational principle ,
   short-plural = s ,
   long-plural = s ,
   short-indefinite = a ,
   long-indefinite = a ,
   tag = abbrev ,
}

\DeclareAcronym{dfvp}{
   short = DFVP ,
   long  = Dirac--Frenkel variational principle ,
}

\DeclareAcronym{ele}{
   short = ELE ,
   long  = Euler--Lagrange equation ,
   short-plural = s ,
   long-plural = s ,
}

\DeclareAcronym{mrcc}{
   short = MRCC ,
   long = multi-reference coupled cluster ,
}

\DeclareAcronym{tdfvci}{
   short = TDFVCI ,
   long = time-dependent full vibrational configuration interaction ,
}

\DeclareAcronym{tdfci}{
   short = TDFCI ,
   long = time-dependent full configuration interaction ,
}

\DeclareAcronym{tdevcc}{
   short = TDEVCC ,
   long  = time-dependent extended vibrational coupled cluster ,
   short-plural = s ,
   long-plural = s ,
   short-indefinite = a ,
   long-indefinite = a ,
   tag = abbrev ,
}

\DeclareAcronym{holc}{
   short = HOLC ,
   long = hybrid optimized and localized vibrational coordinate ,
}

\DeclareAcronym{acf}{
   short = ACF ,
   long = autocorrelation function ,
}

\DeclareAcronym{fwhm}{
   short = FWHM ,
   long  = full width at half maximum ,
   short-plural = s ,
   long-plural = full widths at half maxima ,
   short-indefinite = an ,
   long-indefinite = a ,
   tag = abbrev ,
}

\DeclareAcronym{tdmvcc}{
   short = TDMVCC ,
   long = time-dependent modal vibrational coupled cluster ,
}

\DeclareAcronym{otdmvcc}{
   short = oTDMVCC ,
   long = orthogonal time-dependent modal vibrational coupled cluster ,
}

\DeclareAcronym{rptdmvcc}{
   short = rpTDMVCC ,
   long = restricted polar time-dependent modal vibrational coupled cluster ,
}

\DeclareAcronym{stdmvcc}{
   short = sTDMVCC ,
   long = split time-dependent modal vibrational coupled cluster ,
}

\DeclareAcronym{midas}{
   short = MidasCpp ,
   long = Molecular Interactions{,} Dynamics and Simulations Chemistry Program Package ,
}

\newcommand{\au}{Department of Chemistry, Aarhus University, Langelandsgade 140, 8000 Aarhus C, Denmark}

\begin{document}

\title{A bivariational, stable and convergent hierarchy for time-dependent coupled cluster with adaptive basis sets}

\author{Mads Greisen Højlund}
\email{madsgh@chem.au.dk}
\affiliation{\au}


\author{Ove Christiansen}
\email{ove@chem.au.dk}
\affiliation{\au}

\hypersetup{pdftitle={A bivariational, stable and convergent hierarchy for time-dependent coupled cluster with adaptive basis sets}}
\hypersetup{pdfauthor={M.~G.~Højlund, et al.}}
\hypersetup{bookmarksopen=true}

\date{\today}


\begin{abstract}
    We propose a new formulation of time-dependent \acl{cc} with adaptive basis functions and division of the
one-particle space into active and secondary subspaces.
The formalism is fully bivariational in the sense of a real-valued \acl{tdbvp}
and converges to the complete-active-space solution, a property that is obtained by the use
of biorthogonal basis functions.
A key and distinguishing feature of the theory is that the active bra and ket functions span the
same space by construction.
This ensures numerical stability and is
achieved by employing a split unitary/non-unitary basis
set transformation: The unitary part changes the active space itself, while the non-unitary
part transforms the active basis.
The formulation covers vibrational as well as electron dynamics.
Detailed \aclp{eom} are derived and implemented in the context of vibrational dynamics,
and the numerical behavior is studied and compared to related methods. \end{abstract}

\maketitle

\acresetall

\section{INTRODUCTION} \label{sec:introduction}
Computing the nuclear quantum dynamics of molecules presents a tremendous
challenge for systems with more than a few degrees of freedom. 
One of the most versatile methods that has been developed for this purpose is the \ac{mctdh} method,\cite{meyerMulticonfigurationalTimedependentHartree1990,beckMulticonfigurationTimedependentHartree2000}
which employs a complete wave function expansion inside an active space spanned by a limited
number of adaptive basis functions. The division of the one-particle space into
active and secondary parts allows a compact and highly flexible active space,
something that is a key factor in the success of the method.
An analogous scheme for computing the electron dynamics of molecules
is called \ac{mctdhf}.\cite{zanghelliniMCTDHFApproachMulti2003,katoTimedependentMulticonfigurationTheory2004,nestMulticonfigurationTimedependentHartree2005,caillatCorrelatedMultielectronSystems2005}
While \ac{mctdh} and \ac{mctdhf} constitute important reference approaches in their respective fields, they
involve a computational cost that scales exponentially, 
thus limiting their application to comparatively small systems.
It is therefore highly pertinent to investigate and develop alternative methods that are accurate and achieve polynomial scaling,
a goal that has motivated an immense amount of research in different directions that we will not attempt to survey here.
Instead, we will focus on the \ac{cc} \textit{Ansatz} and, specifically, on the idea
of combining the time-dependent \ac{cc} \textit{Ansatz} with adaptive basis functions.
Adaptive or optimized basis functions have a long history in ground-state electronic \ac{cc} theory,
starting with the \ac{bcc} method,\cite{chilesElectronPairOperator1981,handySizeconsistentBruecknerTheory1989,raghavachariSizeconsistentBruecknerTheory1990,hampelComparisonEfficiencyAccuracy1992}
where the orbitals are determined in such a way that the singles projections vanish.
A closely related approach known as \ac{occ}\cite{purvisFullCoupledclusterSingles1982,scuseriaOptimizationMolecularOrbitals1987,sherrillEnergiesAnalyticGradients1998,krylovSizeconsistentWaveFunctions1998,pedersenGaugeInvariantCoupled1999}
instead minimizes the \ac{cc} energy by using unitary (orthogonal)
orbital transformations. It turns out, however, that \ac{occ} does not reproduce the \ac{fci} limit,\cite{kohnOrbitaloptimizedCoupledclusterTheory2005}
which is of course a disadvantage.
Replacing the unitary (orthogonal) orbital transformation with a non-unitary (non-orthogonal) one
defines the \ac{nocc} method,\cite{pedersenGaugeInvariantCoupled2001}
which does in fact recover the \ac{fci} limit, as shown by Myhre.\cite{myhreDemonstratingThatNonorthogonal2018}

Time-dependent \ac{occ} and \ac{nocc} (\acs{tdocc} and \acs{tdnocc}) were introduced by Pedersen \textit{et al.}\cite{pedersenGaugeInvariantCoupled1999,pedersenGaugeInvariantCoupled2001}
for the purpose of deriving gauge invariant response functions. Concrete working equations for
actually propagating the \ac{tdocc} and \ac{tdnocc} wave functions in time were, however, not provided.
The first example of an explicitly time-dependent \ac{cc} method with dynamical orbitals is thus
Kvaal's \ac{oatdcc}\cite{kvaalInitioQuantumDynamics2012} method from 2012,
which employs biorthogonal (i.e. non-orthogonal) orbitals. Importantly, this method allows
division of the orbital basis and reproduces the \ac{mctdhf} limit correctly.
The \ac{oatdcc} and \ac{tdnocc} \textit{Ansätze} are equivalent if the \ac{oatdcc} orbital space
is not divided into active and secondary parts.

Just as \ac{oatdcc} can be viewed as a combination of the \ac{cc} approach with the basic
idea of \ac{mctdhf}, the \ac{tdmvcc}\cite{madsenTimedependentVibrationalCoupled2020} method
developed in our group combines \ac{vcc} with adaptive basis functions
in the spirit of \ac{mctdh}. We have, however, discovered certain stability issues with the \ac{tdmvcc} approach
that occur specifically when the time-dependent biorthogonal basis set is divided.\cite{hojlundBivariationalTimedependentWave2022}
Our investigation of the problem showed that it is associated with the following fact:
If a biorthogonal basis is divided into active and secondary parts, then there is no guarantee
that the active bra functions and the active ket functions span the same space. In itself, this
is not necessarily a cause for concern, but we have observed that the two spaces tend to drift dangerously far
apart, resulting in serious numerical trouble. In Ref.~\citenum{hojlundBivariationalTimedependentWave2022},
we proposed a so-called restricted polar version of \ac{tdmvcc} (\acs{rptdmvcc}) that forces
the active bra and ket spaces to be the same, while allowing non-orthogonality within the active space.
The \ac{rptdmvcc} model was shown to solve the stability issue without compromising accuracy,
but it should be stressed that the \ac{rptdmvcc} equations are not fully bivariational.
Apart from being somewhat unappealing from a formal standpoint, 
this fact also leads to minor energy non-conservation beyond ordinary numerical noise and integration error.
Even when energy conservation is not of major interest in itself, it serves
as a convenient tool for gauging the integration error (at least for some types of integration schemes)
and is therefore a desirable property in practical computations.

A different approach that cannot suffer from the stability issue described above is to simply
use an orthogonal basis.
Real-time \ac{tdocc} has indeed been introduced by Sato \textit{et al.},\cite{satoCommunicationTimedependentOptimized2018}
and we have recently considered an orthogonal version of \ac{tdmvcc} (\acs{otdmvcc})\cite{hojlundTimedependentCoupledCluster2024a}
analogous to \ac{tdocc}. Our detailed benchmark calculations confirmed numerically that \ac{otdmvcc}
does not converge to \ac{mctdh},
which is certainly a disadvantage.
In concrete terms, the non-convergence means that one cannot generally expect the quality of oTDMVCC
to improve systematically towards the appropriate limit. Although we did not usually observe this deficiency at low
excitation levels,\cite{hojlundTimedependentCoupledCluster2024a} it is surely a problem when high accuracy is required.
Another problem of significant practical 
importance is that the linear equations determining the basis set time evolution
(the so-called constraint equations)
are more involved in
\ac{otdmvcc} than in \ac{tdmvcc} and \ac{rptdmvcc}.\cite{hojlundTimedependentCoupledCluster2024a}

In this paper, we propose a new and fully bivariational formulation of time-dependent \ac{cc} with adaptive basis functions
that combines the following attractive properties: (i) Convergence to the MCTDH/MCTDHF limit, (ii) numerical stability, and (iii)
simple basis set equations. Although we are mainly concerned with the vibrational structure problem, we stress that the formalism
is also applicable to electron dynamics after removing (summations over) mode indices. 
The key idea is to separate the basis set transformation into
two parts: An interspace (active--secondary) transformation
and an intraspace (active--active) transformation. Choosing the former to be unitary and the latter to be non-unitary
results in biorthogonal bra and ket basis functions that span the same space by construction. 
We derive fully bivariational \acp{eom} by applying a real-valued
\ac{tdbvp}\cite{arponenVariationalPrinciplesLinkedcluster1983,kvaalInitioQuantumDynamics2012}
and specialize the formalism to the vibrational dynamics case. 
The resulting method is called split TDMVCC, or \acs{stdmvcc}
for short, since it uses a basis set transformation that is split into
unitary and non-unitary parts.

The paper is organized as follows: Section~\ref{sec:theory} starts with a rather general
explanation of some essential concepts (including adaptive basis functions, biorthogonal bases, and basis set division),
which in turn motivates the introduction of a new \textit{Ansatz} for the basis set transformation. 
Section~\ref{sec:theory} then continues with the details of the theory, including
an outline of the \ac{tdbvp} and derivations of the \acp{eom}. This is followed by a brief description of
our implementation in Sec.~\ref{sec:implementation} and a few numerical examples in Sec.~\ref{sec:results}.
Section~\ref{sec:summary} concludes the paper with a summary of our findings. 

\section{THEORY} \label{sec:theory}

\subsection{Orthogonal and biorthogonal adaptive basis sets} \label{sec:adaptive_basis}
Before deriving any detailed \acp{eom}, we make some general remarks on basis set parameterization with emphasis
on the distinction between orthogonal and biorthogonal bases and on the role of basis set division.
For simplicity, we will refer to an underlying or primitive basis, which is taken to be orthonormal.
The time-dependent basis functions are then expressed (with time-dependent coefficients) 
in terms of the primitive functions. This representation
has been convenient in our previous work, but it is not an integral part of the formalism as such. It is thus fully
possible to derive all equations without reference to an underlying basis, similarly to 
the works of Kvaal\cite{kvaalInitioQuantumDynamics2012} and Sato et al.\cite{satoCommunicationTimedependentOptimized2018}
Such an approach may be more useful if the time-dependent basis is represented on a grid
(Appendix~\ref{appendix:electronic_structure} gives some indications on the connection between
the two approaches).

\begin{figure}[H]
    \centering

    \begin{subfigure}[c]{0.3\textwidth}
        \centering
        \caption{Orthogonal}
        \includegraphics[width=\textwidth]{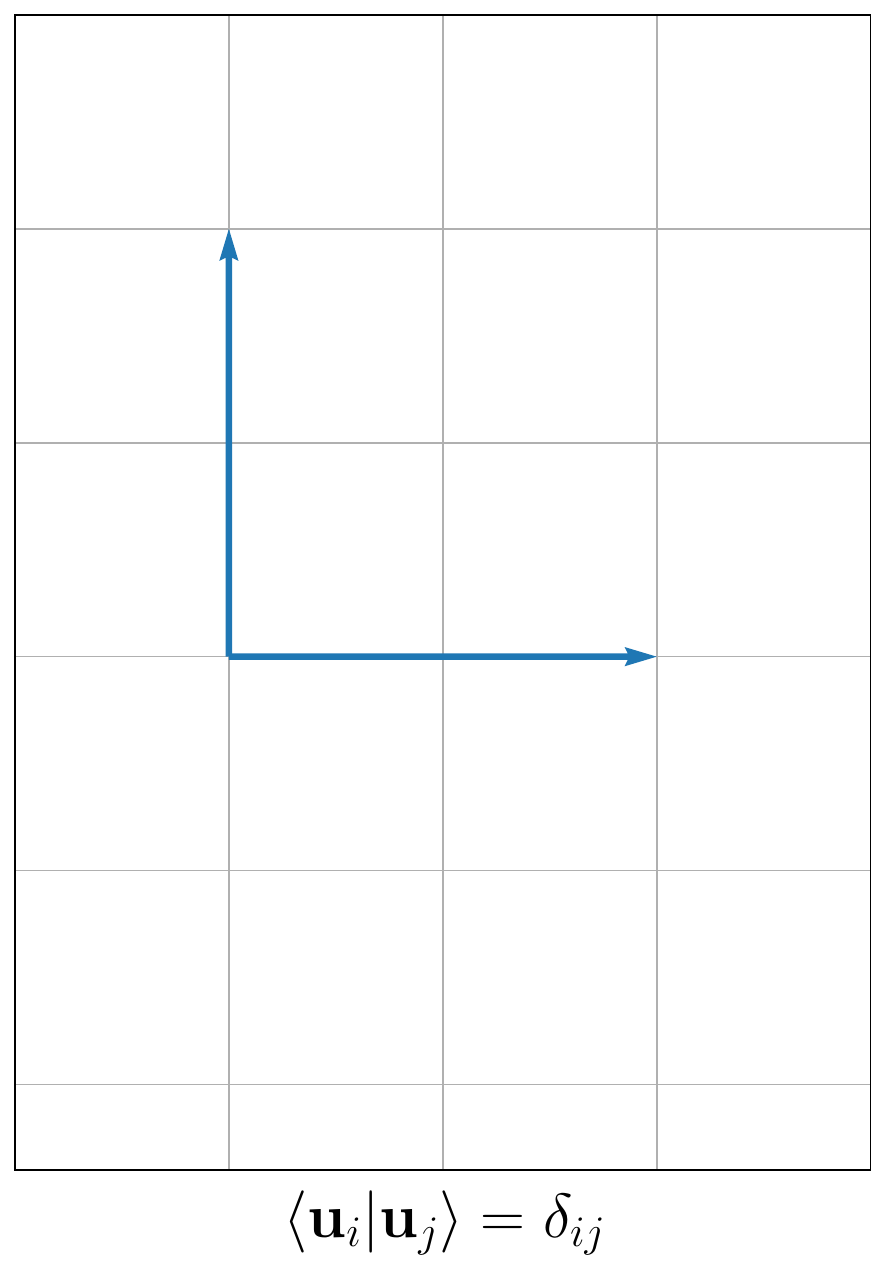}
        \label{fig:2D_basis_ortho}
    \end{subfigure}%
    ~
    \begin{subfigure}[c]{0.3\textwidth}
        \centering
        \caption{Non-orthogonal}
        \includegraphics[width=\textwidth]{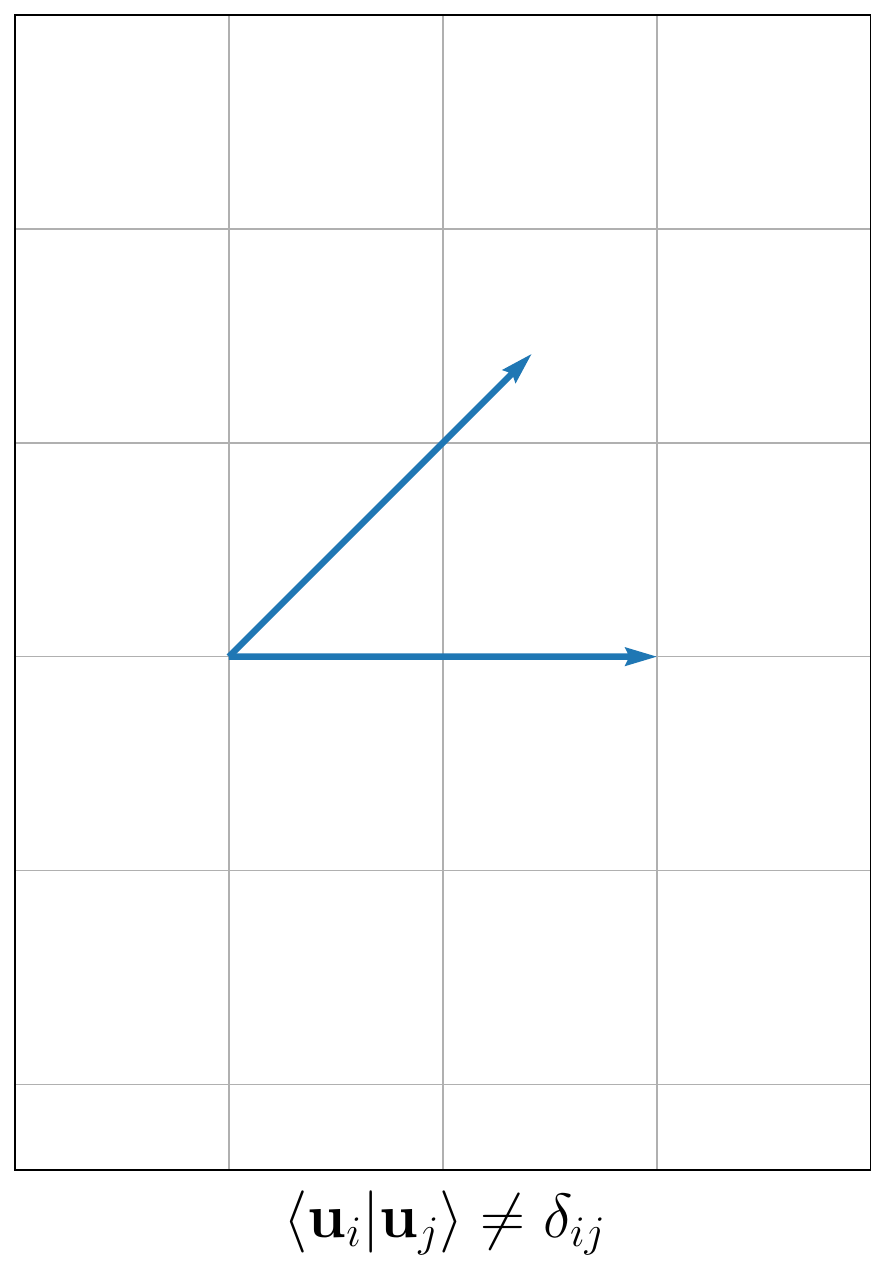}
        \label{fig:2D_basis_nonortho}
    \end{subfigure}%
    ~
    \begin{subfigure}[c]{0.3\textwidth}
        \centering
        \caption{Biorthogonal}
        \includegraphics[width=\textwidth]{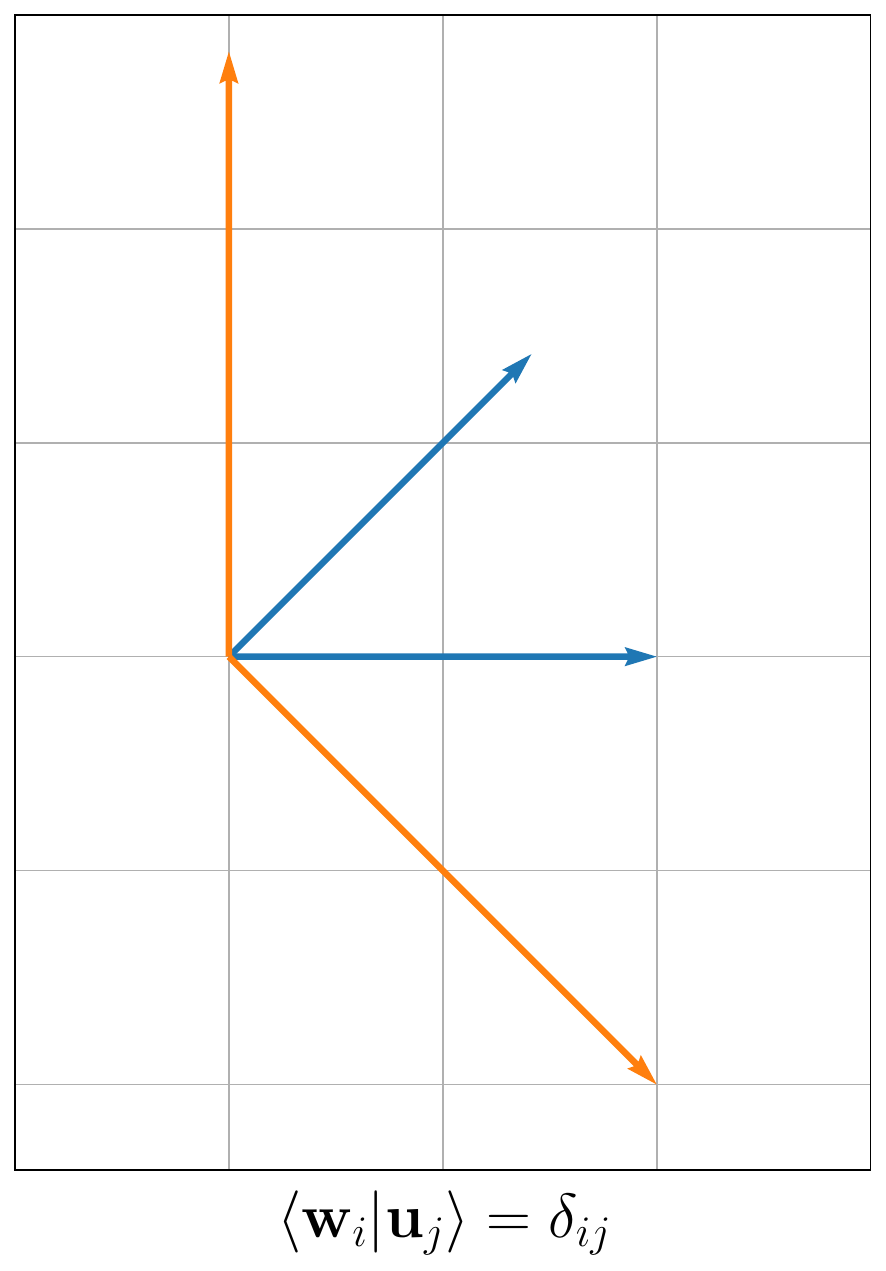}
        \label{fig:2D_basis_biortho}
    \end{subfigure}

    \caption{Three types of bases for the 2D plane. 
    (a) An orthogonal basis (unit overlap). 
    (b) A non-orthogonal basis (non-unit overlap). 
    (c) A biorthogonal basis. The primal/ket basis (blue) and the dual/bra basis (orange) are separate non-orthogonal bases for the plane. 
    Together, they constitute a biorthogonal basis with unit overlap.}
    \label{fig:2D_basis}
\end{figure}

We start by considering separate bra and ket states, $\bra{\psi'}$ and $\ket{\psi}$, that are expressed in the primitive basis.
Any basis set transformation that conserves (bi)orthogonality can now be stated as
\begin{subequations}
    \begin{align}
        \ket{\psi}  &\rightarrow e^{\hat{\kappa}} \ket{\psi}, \\
        \bra{\psi'} &\rightarrow \bra{\psi'}  e^{-\hat{\kappa}},
    \end{align}
\end{subequations}
where $\hat{\kappa}$ is a one-particle operator:
\begin{align}
    \hat{\kappa} = \sum_{m} {\sum_{pq}}^{(m)} \kappaplain{m}{p}{q} \, \Eplain{m}{p}{q}, \quad \Eplain{m}{p}{q} = \crea{m}{p} \anni{m}{q}.
\end{align}
The operators $\crea{m}{p}$ and $\anni{m}{q}$ create and annihilate, respectively, the primitive basis, and we
have used a superscript $(m)$ on the second summation symbol to indicate that there is a separate $p,q$ summation for each vibrational mode
(in particular, the $m$ summation and the superscripted $p,q$ summation do not commute,
which should be kept in mind when reading the paper).
Anti-Hermitian $\hat{\kappa}$ generate unitary transformations that conserve orthogonality, whereas
generic $\hat{\kappa}$ generate invertible transformations that conserve biorthogonality. 
It is a well-known fact\cite{helgakerMolecularElectronicstructureTheory2000} that the 
creators and annihilators transform as 
\begin{subequations}
    \begin{alignat}{4}
        \crea{m}{q} &\rightarrow \creatilde{m}{q} &&= e^{\hat{\kappa}} \, \crea{m}{q} &&e^{-\hat{\kappa}} &&= {\sum_{p}}^{(m)} \crea{m}{p} \exp(\mathbf{K}^m)_{pq},     \label{eq:a_dagger_to_a_dagger_tilde}\\
        \anni{m}{p} &\rightarrow \btilde{m}{p}    &&= e^{\hat{\kappa}} \, \anni{m}{p} &&e^{-\hat{\kappa}} &&= {\sum_{q}}^{(m)} \exp(-\mathbf{K}^m)_{pq} \, \anni{m}{q}, \label{eq:a_to_b_tilde}
    \end{alignat}
\end{subequations}
where $\mathbf{K}^m$ is the matrix containing the $\kappaplain{m}{p}{q}$ parameters.
We note that the transformed annihilators are 
not the adjoint of the transformed creators, except for the special case where $\mathbf{K}^m$ is anti-Hermitian.
The corresponding transformation
for the `first-quantized' basis functions reads as
\begin{subequations}
    \begin{alignat}{2}
        \chiplain{m}{q} &\rightarrow \phitilde{m}{q}      &&= {\sum_{p}}^{(m)} \chiplain{m}{p} \exp(\mathbf{K}^m)_{pq},    \label{eq:chi_to_phi_tilde} \\
        \chiconj{m}{p}  &\rightarrow \phitildeprime{m}{p} &&= {\sum_{q}}^{(m)} \exp(-\mathbf{K}^m)_{pq} \, \chiconj{m}{q}. \label{eq:chi_conj_to_phi_tilde_prime}
    \end{alignat}
\end{subequations}
The transformed functions are generally not each other's complex conjugate
unless $\mathbf{K}^m$ is anti-Hermitian,
and one should therefore think of the primal (or ket) basis [Eq.~\eqref{eq:chi_to_phi_tilde}]
as being separate from the dual (or bra) basis [Eq.~\eqref{eq:chi_conj_to_phi_tilde_prime}].
In the case where all basis functions are active (no basis set division), the difference between
an orthogonal and a biorthogonal basis is easily illustrated by using ordinary vectors in $\mathbb{R}^2$; see Fig.~\ref{fig:2D_basis}.
When the basis is real (as in Fig.~\ref{fig:2D_basis}), the dual basis spans the same space as the primal basis, i.e.
the dual space is identical to the primal space. 
In the complex case, the dual space is exactly the complex conjugate of the primal space.

\begin{figure}[H]
    \centering
    \includegraphics[width=0.70\textwidth]{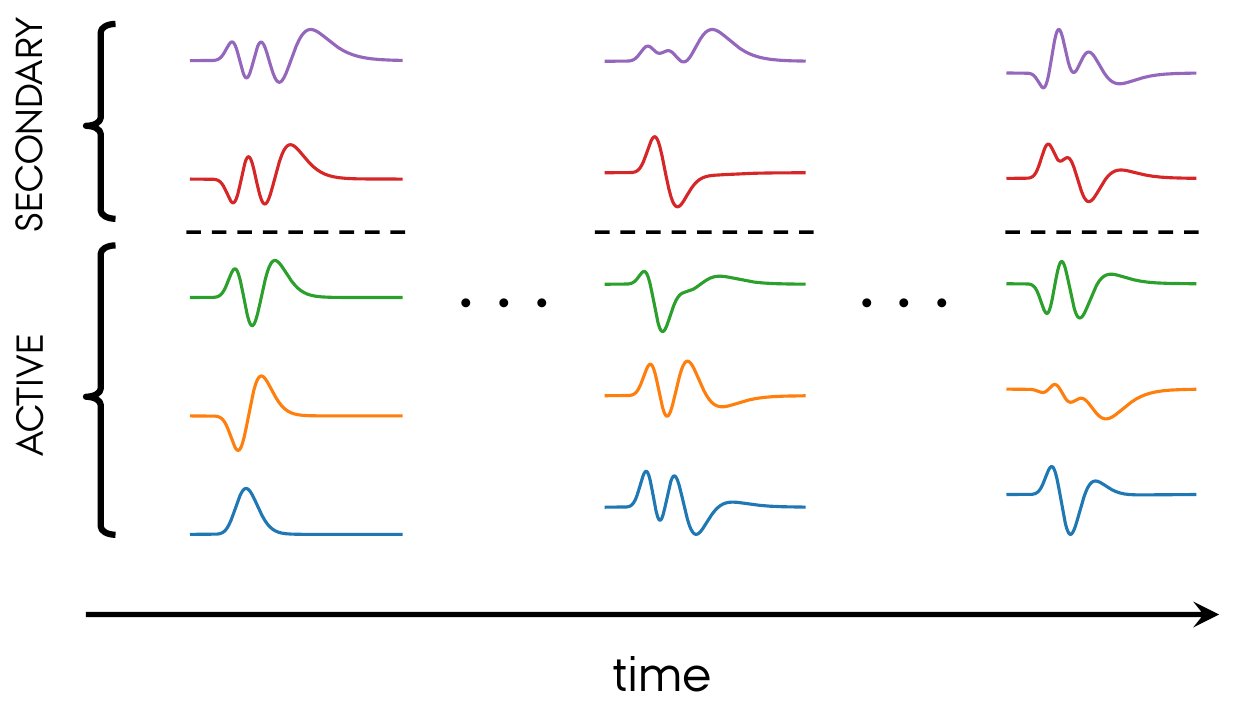}
    \caption{Schematic representation of an adaptive basis that is divided into active and secondary parts. 
    Each vibrational mode has such a basis.}
    \label{fig:td_basis_cartoon}
\end{figure}

\begin{figure}[H]
    \centering

    \begin{subfigure}[c]{0.45\textwidth}
        \centering
        \caption{}
        \includegraphics[width=\textwidth]{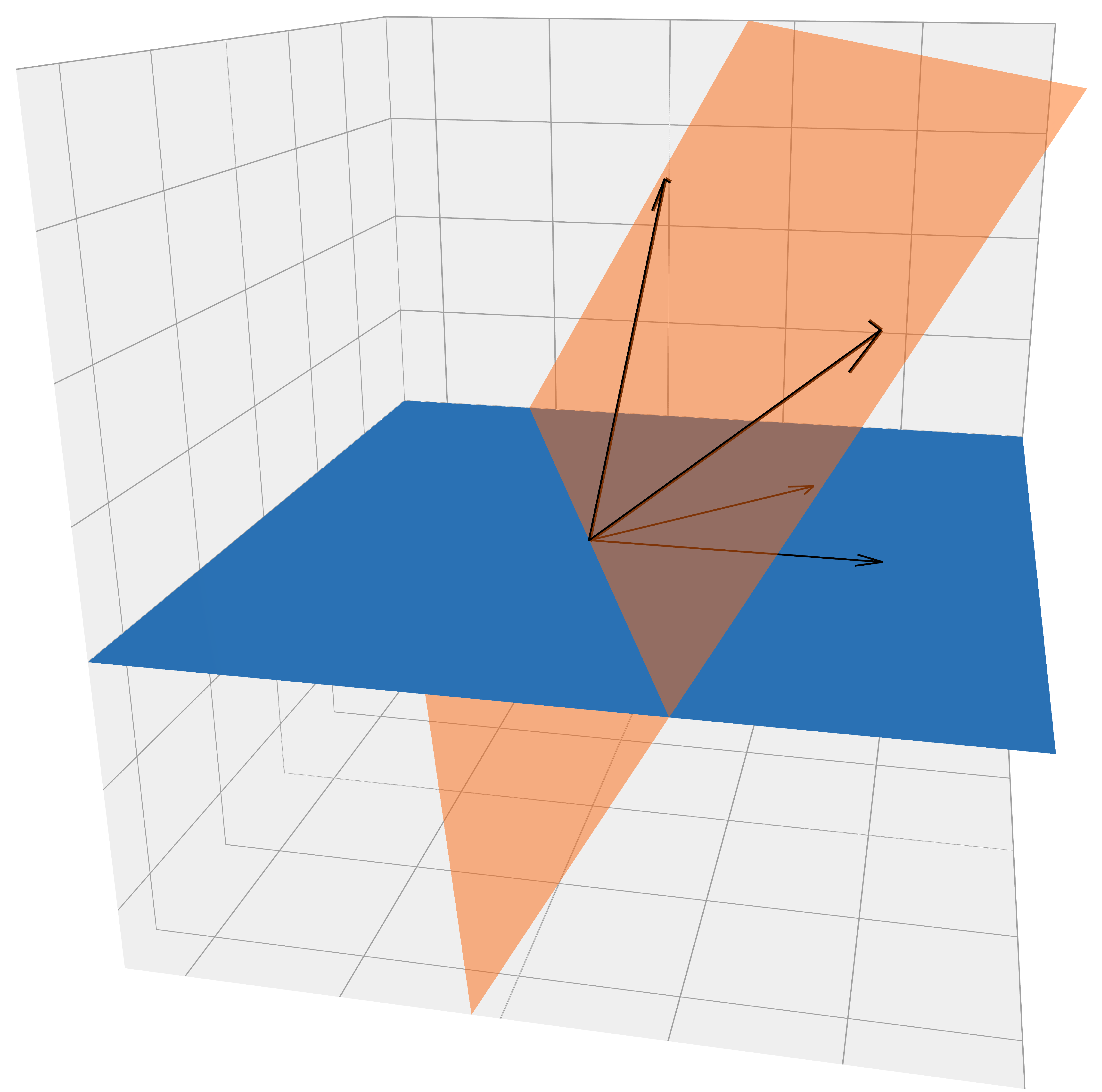}
        \label{fig:3D_basis_from_the_side}
    \end{subfigure}%
    \hspace{1em}
    \begin{subfigure}[c]{0.45\textwidth}
        \centering
        \caption{}
        \includegraphics[width=\textwidth]{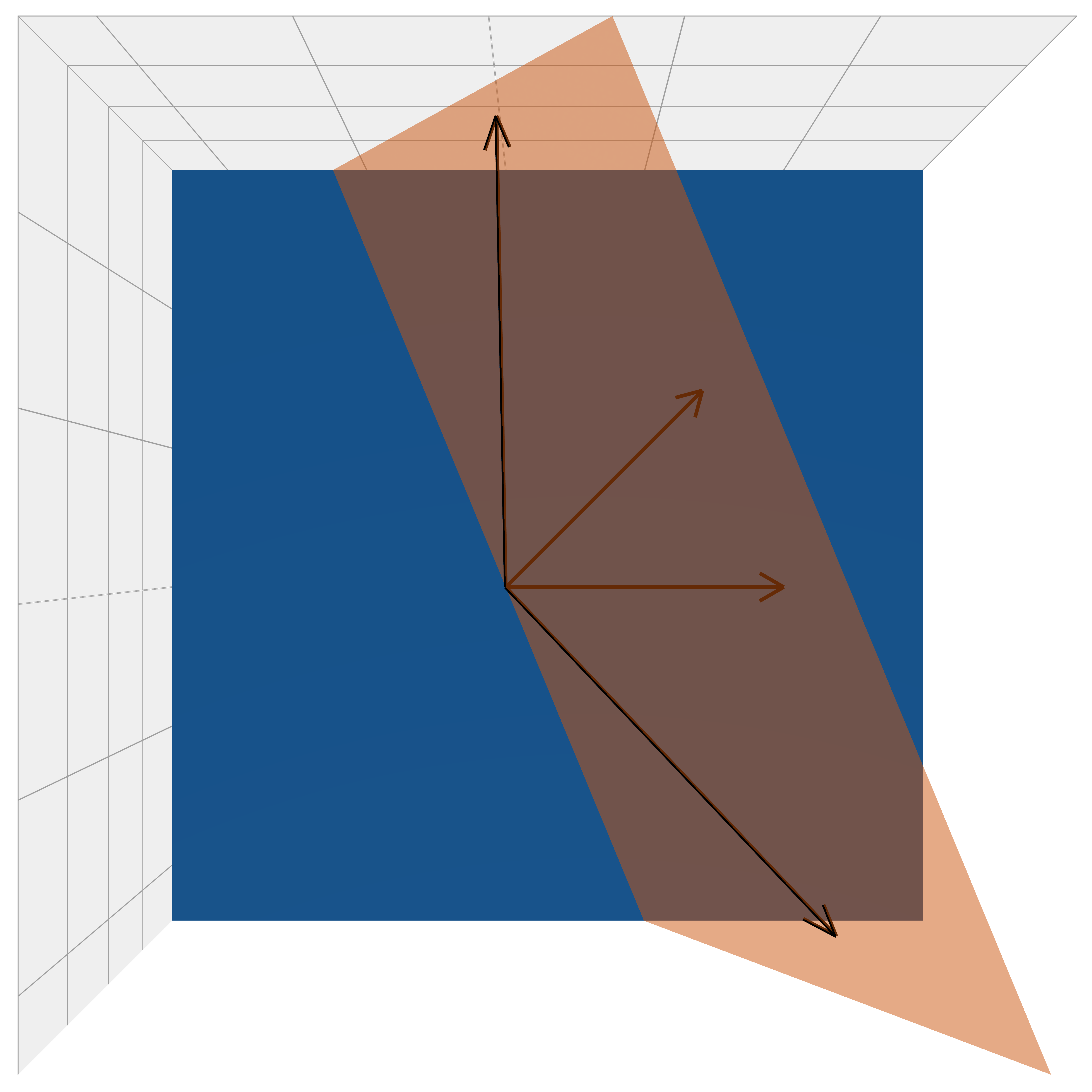}
        \label{fig:3D_basis_from_above}
    \end{subfigure}

    \caption{A biorthogonal set where the primal and dual bases do not span the same 2D planes. 
    The set nevertheless has unit overlap, which can be seen by comparing the right-hand panel to 
    Fig.~\ref{fig:2D_basis_biortho}.}
    \label{fig:3D_basis}
\end{figure}

We now introduce a division of the time-dependent basis into active and secondary parts (see Fig.~\ref{fig:td_basis_cartoon} for a schematic).
The idea is to use only the active functions for constructing the wave function,
while the secondary functions have exactly zero occupation. Such a division enables a
compact and highly flexible active basis, something that has been used to great effect within the
\ac{mctdh}\cite{meyerMulticonfigurationalTimedependentHartree1990, beckMulticonfigurationTimedependentHartree2000}
framework. In the context of \ac{mctdh} (and similar `variational' \textit{Ansätze}) one uses an orthogonal active basis,
so that the active bra space is automatically the conjugate of the active ket space (informally, we will say
that the spaces are the same). This is not so when $\hat{\kappa}$ is a generic one-particle operator, 
as illustrated in Fig.~\ref{fig:3D_basis}. Here, we have constructed a biorthogonal basis for $\mathbb{R}^3$
and selected the two first primal and dual vectors as active basis vectors. The active primal vectors span
the $xy$-plane (blue), while the active dual vectors span a different plane (orange). Nevertheless,
the two sets have unit overlap which can be realized by comparing the right-hand panel in Fig.~\ref{fig:3D_basis}
to Fig.~\ref{fig:2D_basis_biortho}.
This overlap is in fact unchanged when adding arbitrary $z$-components to the dual vectors
(or, more generally, when adding components that are orthogonal to the active primal space).
Requiring unit overlap between the active dual and primal bases is thus a very flexible requirement.
In particular, we note that the dual basis is not uniquely determined by the primal basis when unit overlap is
the only constraint. 

We would like an alternative parameterization where the active dual/bra and primal/ket spaces are
the same, but without restricting the active basis to be orthogonal. 
The first step towards such a parameterization is to understand that the exponential function $e^{\hat{\kappa}}$
has two roles: (i) choosing an active \textit{space} and (ii) choosing a \textit{basis} within that space.
These roles are mixed together in the single exponential formulation, 
but they can be separated by introducing a double or split exponential parameterization:
\begin{subequations}
    \begin{gather}
        e^{\hat{\kappa}} \rightarrow e^{\hat{\sigma}} e^{\hat{\kappa}}, \\
        \hat{\sigma} = \sum_{m} {\sum_{pq}}^{(m)} \sigmaplain{m}{p}{q} \, \Eplain{m}{p}{q}, \quad
        \hat{\kappa} = \sum_{m} {\sum_{pq}}^{(m)} \kappaplain{m}{p}{q} \, \Eplain{m}{p}{q}.
    \end{gather}
\end{subequations}
Here, the first factor ($e^{\hat{\sigma}}$) is an operator that
changes the active \textit{space}, while the second factor ($e^{\hat{\kappa}}$)
changes the active \textit{basis} (the order of the factors will be explained shortly).
Equivalently, one can think of the factors as generating
\textit{interspace} and \textit{intraspace} transformations, i.e. mixing \textit{between} the active and secondary spaces
and mixing \textit{within} the active space.
Such a separation can be realized by requiring a certain type
of structure in the time derivative of the operators $\hat{\sigma}$ and $\hat{\kappa}$. 
The details for the general case (i.e. $\hat{\sigma} \neq 0$ and $\hat{\kappa} \neq 0$)
can be worked out by using the Lie algebraic techniques of Ref.~\citenum{hojlundGeneralExponentialBasis2023a},
but for now we will focus on the simple case where $\hat{\sigma} = 0$ and $\hat{\kappa} = 0$.
At this point, the time derivatives are required to assume a particularly transparent structure, namely
\begin{subequations} \label{eq:sigma_kappa_structure_at_zero}
    \begin{align} 
        \dot{\mathbf{\Sigma}}^m &=
        \left[
        \begin{array}{c | c}
        \mspace{2mu} \mbf{0} \mspace{3mu} {}&{} \mspace{3mu} \bm{*}  \\ \hline
        \mspace{2mu} \bm{*}  \mspace{3mu} {}&{} \mspace{3mu} \mbf{0} \mspace{2mu} {}  
        \end{array} \right], \label{eq:sigma_structure_at_zero} \\
        \dot{\mathbf{K}}^m &=
        \left[
        \begin{array}{c | c}
        \mspace{2mu} \bm{*}   \mspace{3mu} {}&{} \mspace{3mu} \mbf{0} \mspace{2mu} {} \\ \hline
        \mspace{2mu} \mbf{0}  \mspace{3mu} {}&{} \mspace{3mu} \mbf{0} \mspace{2mu} {}  
        \end{array} \right], \label{eq:kappa_structure_at_zero}
    \end{align}
\end{subequations}
where the block structure refers to the
active--secondary division. 
We have introduced the matrices $\mathbf{\Sigma}^m$ and $\mathbf{K}^m$ containing
the $\sigmaplain{m}{p}{q}$ and $\kappaplain{m}{p}{q}$ parameters, respectively,
and used asterisks to indicate non-zero blocks.    
Regardless of the detailed structure, the basic operators transform as
\begin{subequations} \label{eq:double_transformed_operators}
    \begin{alignat}{5}
        \crea{m}{q} 
        &\rightarrow \creabreve{m}{q}
        &&= e^{\hat{\sigma}} e^{\hat{\kappa}} \, \crea{m}{q} &&e^{-\hat{\kappa}} e^{-\hat{\sigma}}
        &&= {\sum_{\bar{p}p}}^{(m)} \crea{m}{\bar{p}}  \exp(+\mathbf{\Sigma}^m)_{\bar{p} p} \exp(+\mathbf{K}^m)_{p q} 
        &&= {\sum_{\bar{p}}}^{(m)}  \crea{m}{\bar{p}} \, \mathcal{U}_{\bar{p}q}^m, \\
        \anni{m}{p} 
        &\rightarrow \bbreve{m}{p}
        &&= e^{\hat{\sigma}} e^{\hat{\kappa}} \, \anni{m}{p} &&e^{-\hat{\kappa}} e^{-\hat{\sigma}}
        &&= {\sum_{\bar{q}q}}^{(m)} \exp(-\mathbf{K}^m)_{p q} \exp(-\mathbf{\Sigma}^m)_{q \bar{q}} \, \anni{m}{\bar{q}}  
        &&= {\sum_{\bar{q}}}^{(m)}  \mathcal{W}_{p\bar{q}}^m \, \anni{m}{\bar{q}},
    \end{alignat}
\end{subequations}
where we have defined matrices $\bm{\mathcal{U}}^m$ and $\bm{\mathcal{W}}^m$ holding
the coefficients of the total transformation for mode $m$ (these matrices are of course the inverse of each other).
It is important to note that the transformations by $\exp(+\mathbf{\Sigma}^m)$ 
and $\exp(-\mathbf{\Sigma}^m)$ are carried out first, contrary
to what one might expect from the ordering of the product $e^{\hat{\sigma}} e^{\hat{\kappa}}$.
The transformation that defines the active space thus precedes the one that defines
a suitable basis within that space.

If we take $\hat{\sigma}$ and $\hat{\kappa}$ to be both anti-Hermitian (or both generic)
we get an overall transformation that is unitary (or non-unitary) and
completely equivalent to a single exponential transformation.
We are, however, free to choose an anti-Hermitian $\hat{\sigma}$ and
a generic $\hat{\kappa}$. This allows unitary transformations of the active space,
and non-unitary transformations within the active space. In particular, the
active bra and ket bases are not necessarily orthogonal, but they span the same space.
The double exponential format is useable as it stands (i.e. one can
propagate the matrices $\mathbf{K}^m$ and $\mathbf{\Sigma}^m$ as the time-dependent basis set
parameters), but we find it convenient to do a slight reformulation.
We are free to absorb the unitary transformation into 
the creators and annihilators and write Eqs.~\eqref{eq:double_transformed_operators}
as
\begin{subequations} \label{eq:double_transformed_operators_alternative}
    \begin{alignat}{3}
        {\sum_{\bar{p}p}}^{(m)} \crea{m}{\bar{p}}  \exp(+\mathbf{\Sigma}^m)_{\bar{p} p} \exp(+\mathbf{K}^m)_{p q}  
        &= {\sum_{p}}^{(m)} \creatilde{m}{p}  \exp(+\mathbf{K}^m)_{p q} 
        &&= e^{\hat{\kappa}} \, \creatilde{m}{q} &&e^{-\hat{\kappa}},
        \\
        {\sum_{\bar{q}q}}^{(m)} \exp(-\mathbf{K}^m)_{p q} \exp(-\mathbf{\Sigma}^m)_{q \bar{q}} \, \anni{m}{\bar{q}}  
        &= {\sum_{q}}^{(m)} \exp(-\mathbf{K}^m)_{p q} \, \annitilde{m}{q} 
        &&= e^{\hat{\kappa}} \, \annitilde{m}{p} &&e^{-\hat{\kappa}}.
    \end{alignat}
\end{subequations}
Although we have not changed the notation, the $\hat{\kappa}$ operator
is now defined with respect to the moving creators $\creatilde{m}{q}$:
\begin{align} \label{eq:moving_creators}
    \hat{\kappa} = \sum_{m} {\sum_{pq}}^{(m)} \kappaplain{m}{p}{q} \Etilde{m}{p}{q}, \quad \Etilde{m}{p}{q} = \creatilde{m}{p} \annitilde{m}{q}.
\end{align}
These creators are themselves given by
\begin{align}
    \creatilde{m}{p} 
    = {\sum_{\bar{p}}}^{(m)} \crea{m}{\bar{p}}  \exp(\mathbf{\Sigma}^m)_{\bar{p} p}
    = {\sum_{\bar{p}}}^{(m)} \crea{m}{\bar{p}}  \vplain{m}{\bar{p}}{p},
\end{align}
where the matrix $\mathbf{V}^m$ is unitary and can be propagated in time
rather than $\mathbf{\Sigma}^m$. The linear parameterization in terms of $\mathbf{V}^m$ 
is (in some ways) simpler than the exponential parameterization in terms of $\mathbf{\Sigma}^m$,
but it requires the use of constraints, a point that we will return to later.
The main motivation for using a linear-type parameterization for the unitary transformation
is that it will allow us to introduce a secondary-space projection, which is convenient
if the secondary space is large. 

It is possible to also
absorb the non-unitary transformation
into the creators/annihilators from the outset, 
which would correspond to working explicitly in the breve basis of Eqs.~\eqref{eq:double_transformed_operators}
with the matrices $\bm{\mathcal{U}}^m$ and $\bm{\mathcal{W}}^m$ as the time-dependent parameters.
This leads, however,
to complicated constraints and less transparent derivations (as far as we can tell).
We will therefore
keep the exponential formulation for the non-unitary transformation for the time being
and only describe an optional linearization at the end. In conclusion, we continue with
a parameterization like
\begin{subequations}
\begin{align}
    \ket{\Psi}  &= e^{\hat{\kappa}} \ket{\psi}, \\
    \bra{\Psi'} &= \bra{\psi'} e^{\hat{\kappa}},
\end{align}
\end{subequations}
where $\ket{\psi}$, $\bra{\psi'}$ and $\hat{\kappa}$ are expressed in terms of the
moving creators of Eq.~\eqref{eq:moving_creators}.
With this choice, the $\mathbf{K}^m$ matrices always have the overall structure
given by Eq.~\eqref{eq:kappa_structure_at_zero} and the total basis set transformation
is given by the matrices
\begin{subequations}
\begin{gather}
    \bm{\mathcal{U}}^m 
    = \mathbf{V}^m \exp(+\mathbf{K}^m)
    =
    \left[
	\begin{array}{c | c}
        \mbf{V}_{\!\!\subA}^m {\,} & {\,} \mbf{V}_{\!\subS}^m \Tstrut
	\end{array} \right]
    \left[
	\begin{array}{c | c}
	{\mbf{U}^m_{\!\subA\subA}} {\,} & {\,} {\mbf{0}} \Tstrut
	\\ \hline
	{\mbf{0}} {\,} & {\,} \mspace{6mu} {\mbf{1}} {\mspace{6mu}}  \Tstrut
	\end{array} \right],
    \\
    \bm{\mathcal{W}}^m 
    = \exp(-\mathbf{K}^m) \, \mathbf{V}^{m\dagger}
    =
    \left[
	\begin{array}{c | c}
	{\mbf{W}^m_{\!\!\subA\subA}} {\,} & {\,} {\mbf{0}} \Tstrut
	\\ \hline
	{\mbf{0}} {\,} & {\,} \mspace{6mu} {\mbf{1}} {\mspace{6mu}}  \Tstrut
	\end{array} \right]
    \left[
	\begin{array}{c}
        \mbf{V}_{\!\!\subA}^{m\dagger} \\ \hline
        \mbf{V}_{\!\subS}^{m\dagger} 
	\end{array} \right],
\end{gather}
\end{subequations}
where we have defined $\mathbf{U}^m = \exp(\mathbf{K}^m)$ and $\mathbf{W}^m = \exp(-\mathbf{K}^m)$
and used the block structure of $\mathbf{K}^m$.
In particular, the active blocks of $\bm{\mathcal{U}}^m $ and $\bm{\mathcal{W}}^m$ are
\begin{subequations}
\begin{align}
    \bm{\mathcal{U}}^m_{\!\subA}   &= \mathbf{V}^m_{\!\!\subA} \mathbf{U}^m_{\!\subA\subA}, \\
    \bm{\mathcal{W}}^m_{\!\!\subA} &= \mathbf{W}^m_{\!\!\subA\subA} \mathbf{V}^{m\dagger}_{\!\!\subA}.
\end{align}
\end{subequations}
These matrices are exactly each other's Moore--Penrose inverses,\cite{weissteinMoorePenroseMatrixInverse2008}
which is another way of stating the fact that the active bra and ket spaces are the same,
even though the active basis is not orthogonal. The Moore--Penrose inverse is unique whenever it exists, 
so the active bra basis is uniquely defined by the active ket basis (and vice versa).

We have obtained the Moore--Penrose structure in a particular way, namely by
separating interspace and intraspace transformations, but this is not the only way.
Returning to the double exponential format at the point $\hat{\sigma} = \hat{\kappa} = 0$,
we could also require 
\begin{subequations} \label{eq:sigma_kappa_structure_at_zero_polar}
    \begin{align} 
        \dot{\mathbf{\Sigma}}^m &=
        \left[
        \begin{array}{c | c}
        \mspace{2mu} \bm{*}  \mspace{3mu} {}&{} \mspace{3mu} \bm{*}  \\ \hline
        \mspace{2mu} \bm{*}  \mspace{3mu} {}&{} \mspace{3mu} \mbf{0} \mspace{2mu} {}  
        \end{array} \right], \label{eq:sigma_structure_at_zero_polar} \\
        \dot{\mathbf{K}}^m &=
        \left[
        \begin{array}{c | c}
        \mspace{2mu} \bm{*}   \mspace{3mu} {}&{} \mspace{3mu} \mbf{0} \mspace{2mu} {} \\ \hline
        \mspace{2mu} \mbf{0}  \mspace{3mu} {}&{} \mspace{3mu} \mbf{0} \mspace{2mu} {}  
        \end{array} \right], \label{eq:kappa_structure_at_zero_polar}
    \end{align}
\end{subequations}
with $\dot{\mathbf{\Sigma}}^m$ anti-Hermitian and $\dot{\mathbf{K}}^m$ Hermitian.
This leads to a polar-type decomposition\cite{hallLieGroupsLie2015} of the overall basis set transformation
with $e^{\hat{\sigma}}$ unitary and $e^{\hat{\kappa}}$ positive definite,
and where the latter factor is restricted to acting within the active space.
We arrived at such a `restricted polar' parameterization from a different starting point in Ref.~\citenum{hojlundBivariationalTimedependentWave2022},
but the \acp{eom} obtained were not fully bivariational. Although we did not observe
any deterioration of accuracy, we did see some energy non-conversation as a symptom
of the non-bivariational nature of the theory. It should be possible to derive 
a fully bivariational version of this theory, but we believe it would be
rather complicated compared to the approach of separating interspace and intraspace transformations.

The idea of a double exponential basis set transformation has in fact been considered
in ground state electronic structure theory. Köhn and Olsen\cite{kohnOrbitaloptimizedCoupledclusterTheory2005} 
considered a `modified OCC' method defined by a (bi)variationally optimized, unitary interspace transformation
and a Brueckner-type intraspace transformation determined by projection. This method recovers the \ac{fci} 
limit for ground state energies, but inherits the unphysical second-order poles in the
\ac{bcc} response function.\cite{aigaFrequencydependentHyperpolarizabilitiesBrueckner1994,kochBruecknerCoupledCluster1994} 
It is also not clear to us how one would actually propagate a Brueckner-type
transformation in time. Consequently, we do not consider the `modified \ac{occ}' \textit{Ansatz}
as an attractive choice for our purposes. We also mention the work of Olsen\cite{olsenNovelMethodsConfiguration2015} on \acs{ci}-type
expansions with non-orthogonal orbitals that are optimized using a polar-type product $e^{\hat{\sigma}} e^{\hat{\kappa}}$, where
$\hat{\sigma}$ is anti-Hermitian (anti-symmetric) and $\hat{\kappa}$ is Hermitian (symmetric). Although this approach is highly
relevant for the present discussion, it is not directly applicable in the context of wave functions that
are explicitly time-dependent, complex-valued, and bivariational.


\subsection{The Hamiltonian in second quantization} \label{sec:sq}
The first-quantized Hamiltonian can be written as
\begin{align}
    \hat{H} = \sum_{\mathbf{m}} \hat{H}^{\mathbf{m}},
\end{align}
where $\mathbf{m}$ is a \ac{mc}, i.e. a set of modes.
The notation means that a given term $\hat{H}^{\mathbf{m}}$ acts
only on the modes $\mathbf{m}$. In the second-quantized formalism,\cite{christiansenSecondQuantizationFormulation2004}
the Hamiltonian with respect to the primitive basis reads
\begin{align} \label{eq:H_primitive_basis}
    H = \sum_{\mathbf{m}} {\sum_{\bm{\alpha} \bm{\beta}}}^{(\mathbf{m})} H^{\mathbf{m}}_{\bm{\alpha} \bm{\beta}} \Eplain{\mathbf{m}}{\bm{\alpha}}{\bm{\beta}}, 
    \quad \Eplain{\mathbf{m}}{\bm{\alpha}}{\bm{\beta}} = \prod_{m \in \mathbf{m}} \Eplain{m}{\alpha^m}{\beta^m},
\end{align}
with integrals
\begin{align}
    H^{\mathbf{m}}_{\bm{\alpha} \bm{\beta}} = \Big\langle \prod_{m \in \mathbf{m}} \chiplain{m}{\alpha^m} \Big| \hat{H}^{\mathbf{m}} \Big| \prod_{m \in \mathbf{m}} \chiplain{m}{\beta^m} \Big\rangle.
\end{align}
Here, $\bm{\alpha}$ and $\bm{\beta}$ are vectors of primitive indices, which are henceforth represented by lowercase Greek letters.
Each operator $\hat{H}^{\mathbf{m}}$ is typically represented as a sum of products of one-mode operators, in which case the integrals factor into products of one-mode integrals.
In the basis defined by the moving creators of Eq.~\eqref{eq:moving_creators}, the operator looks like
\begin{gather} \label{eq:H_moving_basis}
    H = \sum_{\mathbf{m}} {\sum_{\mathbf{p} \mathbf{q}}}^{(\mathbf{m})} \tilde{H}^{\mathbf{m}}_{\mathbf{p} \mathbf{q}} \Etilde{\mathbf{m}}{\mathbf{p}}{\mathbf{q}}, \\
    \tilde{H}^{\mathbf{m}}_{\mathbf{p} \mathbf{q}} 
    = {\sum_{\bm{\alpha} \bm{\beta}}}^{(\mathbf{m})} 
    \left( \prod_{m \in \mathbf{m}} \vplainconj{m}{\alpha^m}{p^m}  \right)  H^{\mathbf{m}}_{\bm{\alpha} \bm{\beta}} \left( \prod_{m \in \mathbf{m}} \vplain{m}{\beta^m}{q^m}  \right),
\end{gather}
but we remark that Eqs.~\eqref{eq:H_primitive_basis} and \eqref{eq:H_moving_basis} are strictly equal. We will also encounter
the similarity transformed Hamiltonian
\begin{align} \label{eq:H_bar_def}
    \bar{H} 
    = e^{-\hat{\kappa}} H e^{\hat{\kappa}}
    = \sum_{\mathbf{m}} {\sum_{\mathbf{p} \mathbf{q}}}^{(\mathbf{m})} \breve{H}^{\mathbf{m}}_{\mathbf{p} \mathbf{q}} \Etilde{\mathbf{m}}{\mathbf{p}}{\mathbf{q}}.
\end{align}
The creators and annihilators are still in the basis of Eq.~\eqref{eq:moving_creators}, but the integrals have been transformed once more:
\begin{align}
    \breve{H}^{\mathbf{m}}_{\mathbf{p} \mathbf{q}}
    &= {\sum_{\bar{\mathbf{p}} \bar{\mathbf{q}}}}^{(\mathbf{m})} 
    \left( \prod_{m \in \mathbf{m}} \exp(-\mathbf{K}^m)_{p^m \bar{p}^m}  \right)  \tilde{H}^{\mathbf{m}}_{\bar{\mathbf{p}} \bar{\mathbf{q}}} \left( \prod_{m \in \mathbf{m}} \exp(\mathbf{K})_{\bar{q}^m q^m}  \right) \nn
    &= {\sum_{\bm{\alpha} \bm{\beta}}}^{(\mathbf{m})} 
    \left( \prod_{m \in \mathbf{m}} \mathcal{W}^m_{p^m \alpha^m}  \right)  H^{\mathbf{m}}_{\bm{\alpha} \bm{\beta}}                    \left( \prod_{m \in \mathbf{m}} \mathcal{U}^m_{\beta^m q^m}  \right).
\end{align}
 
\subsection{The time-dependent bivariational principle} \label{sec:tdbvp}
The \ac{tdbvp} in its original form\cite{arponenVariationalPrinciplesLinkedcluster1983,kvaalInitioQuantumDynamics2012}
considers a complex-valued Lagrangian,
\begin{align} \label{eq:complex_lagrangian_general}
    \quad \mathcal{L} = \elm{\Psi'}{(i \partial_t - H)}{\Psi},
\end{align}
and the corresponding complex-valued action,
\begin{align}
    \mathcal{S} = \int_{t_0}^{t_1} \mathcal{L} \dd{t}.
\end{align}
The bra and ket states are taken to be independent.
Requiring the action to be stationary ($\delta \mathcal{S} = 0$) with respect to arbitrary
and independent variations that vanish at the end points is equivalent
to the Schrödinger equation and its dual:
\begin{subequations}
\begin{align}
     i \ket{\dot{\Psi}}  = H \ket{\Psi}, \\
    -i \bra{\dot{\Psi}'} = \bra{\Psi'} H.
\end{align}
\end{subequations}
In practice, the bra and ket states are parameterized in terms of a set of complex
parameters $z_i$ and the stationary condition is then equivalent to a set
of \acp{ele},
\begin{align}
    0 &= \pdv{\mathcal{L}}{z_i}   - \dv{t} \pdv{\mathcal{L}}{\dot{z}_i}.
\end{align}
These equations are consistent as long as the Lagrangian is holomorphic (complex analytic) in
the parameters (in particular, complex conjugation must not appear in $\mathcal{L}$).
This subtle point is mentioned by Kvaal\cite{kvaalInitioQuantumDynamics2012} and discussed
in some detail in Ref.~\citenum{hojlundTimedependentCoupledCluster2024a} and its Appendix A. Although it may seem like a technicality, it
has some important consequences for the kinds of parameterizations that can be permitted.
As an example, the complex-valued Lagrangian rules out the use of an orthogonal basis 
(in that case the bra basis is simply the complex conjugate of the ket basis).

Instead, we may consider a manifestly real Lagrangian,
\begin{align} \label{eq:real_lagrangian}
    \bar{\mathcal{L}} = \mathrm{Re}(\mathcal{L}) = \tfrac{1}{2} (\mathcal{L} + \mathcal{L}^*),
\end{align}
where we allow the parameterization of the bra and ket states to be non-holomorphic, 
i.e. we allow the states to depend on a set of parameters $\mbf{z}$ as well
as the complex conjugate parameters $\mathbf{z}^*$. 
The total set of independent variational parameters is thus $\{ \mbf{z}, \mbf{z}^* \}$
(it is also possible to use $\mathrm{Re}(\mbf{z})$ and $\mathrm{Im}(\mbf{z})$ as the independent parameters, 
but this is somewhat inconvenient in practical derivations).
The evolution of the parameters is determined by making the real action functional
$\bar{\mathcal{S}} = \mathrm{Re}(\mathcal{S})$ stationary, which
is equivalent to a set of \ac{ele} for the parameters $\mbf{z}$ and $\mbf{z}^*$ separately, i.e. 
\begin{subequations} \label{eq:eles_direct_and_conj_params}
    \begin{align} 
        0 &= \pdv{\bar{\mathcal{L}}}{z_i}   - \dv{t} \pdv{\bar{\mathcal{L}}}{\dot{z}_i}  , \label{eq:eles_direct_and_conj_params_a} \\
        0 &= \pdv{\bar{\mathcal{L}}}{z_i^*} - \dv{t} \pdv{\bar{\mathcal{L}}}{\dot{z}_i^*}. \label{eq:eles_direct_and_conj_params_b}
    \end{align}
\end{subequations}
These equations are each other's complex conjugate, since the Lagrangian is real. It is thus sufficient to solve
one of them.

We note that if $\mathcal{L}$ depends on a certain parameter $z_k$, but not on the complex conjugate $z_k^*$, then
the real-valued \ac{tdbvp} reduces to the complex-valued
\ac{tdbvp} for that particular parameter:
\begin{align}
    0 
    &= \pdv{\bar{\mathcal{L}}}{z_k}   - \dv{t} \pdv{\bar{\mathcal{L}}}{\dot{z}_k} \nn 
    &= \frac{1}{2} \left( \pdv{\mathcal{L}}{z_k}   - \dv{t} \pdv{\mathcal{L}}{\dot{z}_k} \right)
    +  \frac{1}{2} \left( \pdv{\mathcal{L}}{z_k^*}   - \dv{t} \pdv{\mathcal{L}}{\dot{z}_k^*} \right)^{\!*} \nn 
    &= \frac{1}{2} \left( \pdv{\mathcal{L}}{z_k}   - \dv{t} \pdv{\mathcal{L}}{\dot{z}_k} \right).
\end{align}

\subsection{Parameterization and constraints} \label{sec:parameterization}
As explained in Sec.~\ref{sec:adaptive_basis}, we consider a moving orthonormal
basis or, equivalently, a set of moving creators
\begin{align}
    \creatilde{m}{p} = {\sum_{\alpha}}^{(m)} \crea{m}{\alpha} \vplain{m}{\alpha}{p}.
\end{align}
At this point we consider all basis functions (active and secondary) explicitly, so that
the matrix $\mathbf{V}^m$ containing the expansion coefficients is a square, unitary matrix.

The time-dependent basis must be orthonormal at any given time,
which is a constraint on the basis set. In addition, the
time derivative of this constraint must vanish at any given time
(this is a so-called consistency condition\cite{diracGeneralizedHamiltonianDynamics1950,ohtaTimedependentVariationalPrinciple2000,ohtaTimedependentVariationalPrinciple2004}).
Using matrix notation, the constraints and consistency conditions read
\begin{subequations}
\begin{gather}
    \mathbf{V}^{m \dagger} \, \mathbf{V}^m = \mathbf{1}^m, \\ 
    \mathbf{0} = \dot{\mathbf{V}}^{m \dagger} \, \mathbf{V}^m
    + \mathbf{V}^{m \dagger} \, \dot{\mathbf{V}}^m.
\end{gather}
\end{subequations}
The latter equation holds trivially if
\begin{subequations} \label{eq:consistency}
    \begin{align}
        \dot{\mathbf{V}}^{m \dagger} \, \mathbf{V}^m &= +i \mathbf{G}^m, \label{eq:consistency_a} \\
        \mathbf{V}^{m \dagger} \, \dot{\mathbf{V}}^m &= -i \mathbf{G}^m, \label{eq:consistency_b}
    \end{align}
\end{subequations}
where $\mathbf{G}^m$ is a Hermitian but otherwise arbitrary matrix, which
we call the constraint matrix. Using Eq.~\eqref{eq:consistency_b}
and the fact that $\mathbf{V}^m$ is unitary we see that
\begin{align} \label{eq:V_dot_general}
    i \dot{\mathbf{V}}^m = \mathbf{V}^m \mathbf{G}^m,
\end{align}
which means that the constraint matrix acts as a generator for the time evolution of $\mathbf{V}^m$.

It is convenient at this point to divide the time-dependent basis
into an active basis indexed by $t$, $u$, $v$, $w$
and a secondary basis indexed by $x$, $y$.
We use $\alpha$, $\beta$ to denote primitive indices, while $p$, $q$, $r$, $s$
denote generic time-dependent indices. The division of the basis induces
a block structure in $\mathbf{V}^m $ and $\mathbf{G}^m$:
\begin{subequations}
    \begin{gather}
        \mbf{V}^m =
        \left[
        \begin{array}{c | c} 
        {\mbf{V}^m_{\!\!\subA} \,} & \, \mbf{V}^m_{\!\subS}  \bottomstrut{-1.0ex}
        \end{array}\right], \\
        \mbf{G}^m =
        \left[
        \begin{array}{c | c}
        {\mbf{G}^m_{\subA\subA}}  & {\mbf{G}^m_{\subA\subS}} \Tstrut
        \\ \hline
        {\mbf{G}^m_{\subS\subA}}  & {\mbf{G}^m_{\subS\subS}}   \Tstrut
        \end{array} \right].
    \end{gather}
\end{subequations}
Right-hand subscripts $\textsc{a}$ and $\textsc{s}$ are used to indicate the active and secondary
blocks of matrices that have one primitive and one time-dependent index.
Matrices with two time-dependent indices have four blocks:
(i) active--active ($\textsc{a}\textsc{a}$); (ii) active--secondary ($\textsc{a}\textsc{s}$); (iii)
secondary--active ($\textsc{s}\textsc{a}$); and (iv) secondary--secondary ($\textsc{s}\textsc{s}$).

In order to ensure that $\mathbf{G}^m$ generates only interspace transformations,
we enforce the block structure of Eq.~\eqref{eq:sigma_structure_at_zero}, i.e. we set
\begin{align} \label{eq:G_structure}
    \mbf{G}^m \equiv
    \left[
    \begin{array}{c | c}
        \mbf{0} & {\mbf{G}^m_{\subA\subS}} \Tstrut
    \\ \hline
    {\mbf{G}^m_{\subS\subA}}  & \mbf{0}   \Tstrut
    \end{array} \right].
\end{align}
The $\mbf{G}^m_{\subS\subS}$ block is always redundant and can safely be set to zero,
whereas the $\mbf{G}^m_{\subA\subA}$ block would ordinarily not be redundant. 
It is, however, not needed in our case since we have explicitly introduced a separate parameterization
for the intraspace transformations. In conclusion, we only retain the 
$\mbf{G}^m_{\subS\subA}$ and $\mbf{G}^m_{\subA\subS} = (\mbf{G}^m_{\subS\subA})^{\dagger}$ blocks, which
mix the active and secondary spaces.
It is important to note that this particular \textit{Ansatz} for $\mbf{G}^m$ generates unitary transformations
that change the active space without determining an appropriate
active basis. That is to say that the columns of $\mbf{V}^m_{\!\!\subA}$ span
the active space, but they are generally not good basis vectors in themselves.


To summarize: We will consider rather general bra and ket parameterizations of the form
\begin{subequations}
    \begin{align}
        \ket{\Psi  (\bm{\alpha}, \bm{\kappa}, \mbf{V})  } &= e^{\hat{\kappa}} \ket{\psi (\bm{\alpha}, \mbf{V}) }, \\
        \bra{\Psi' (\bm{\alpha}, \bm{\kappa}, \mbf{V}^*)} &= \bra{\psi' (\bm{\alpha}, \mbf{V}^*) } e^{-\hat{\kappa}},
    \end{align}
\end{subequations}
with 
\begin{align} \label{eq:kappa_def}
    \hat{\kappa} = \sum_{m} {\sum_{tu}}^{(m)} \kappaplain{m}{t}{u} \Etilde{m}{t}{u}, \quad \Etilde{m}{t}{u} = \creatilde{m}{t} \annitilde{m}{u}.
\end{align}
The $\kappaplain{m}{t}{u}$ parameters are collected in vectors $\bm{\kappa}^m_{\subA\subA}$ or matrices $\mbf{K}^m_{\subA\subA}$ depending
on context (the vector $\bm{\kappa}_{\subA\subA}$ contains all $\kappaplain{m}{t}{u}$ for all modes).
$\bm{\alpha}$ is a vector of complex configurational or correlation parameters (e.g. \ac{cc} amplitudes) and
$\mbf{V} = \{ \mbf{V}^m \}$ denotes the basis set coefficients. It is important to note
that the ket state depends on the coefficients themselves, while the bra state depends
on the complex conjugate coefficients, $\mbf{V}^*$.  
\subsection{Equations of motion} \label{sec:eom}
\subsubsection{Lagrangian}
In order to compute the Lagrangian, Eq.~\eqref{eq:real_lagrangian},
we will start by considering the action of the time-derivative on
a single Hartree product:
\begin{align}
    \ket{\Phi_\mbf{s}} = \prod_{m=1}^{M} \creatilde{m}{s^m} \ket{\mrm{vac}}
\end{align}
It is easy to see by applying the product rule that the time derivative
acts as a one-particle operator on a single Hartree product (the same holds
for a single Slater determinant). In fact, one can show (see Ref.~\citenum{hojlundBivariationalTimedependentWave2022})
that
\begin{align}
    \elm{\Phi_\mbf{r}}{\partial_t}{\Phi_\mbf{s}} = \elm{\Phi_\mbf{r}}{(-iG)}{\Phi_\mbf{s}}
\end{align}
where the so-called constraint operator,
\begin{align}
    G = \sum_{m} {\sum_{pq}}^{(m)} \gplain{m}{p}{q} \, \Etilde{m}{p}{q},
\end{align}
is constructed from the constraint matrix elements. Next, we introduce an active-space
resolution of identity and write the bra and ket as
\begin{subequations} \label{eq:states_as_linear_expansion}
    \begin{alignat}{2}
        \ket{\Psi}  
        &= \sum_{\mu} \ket{\mu} \2\2 \braket{\mu|\Psi}
        &&\equiv  \sum_{\mu} c_{\mu}(\bm{\alpha}, \bm{\kappa})  \ket{\mu(\mbf{V})}, \label{eq:ket_as_linear_expansion}\\
        \bra{\Psi'} 
        &= \sum_{\mu}   \braket{\Psi' | \mu} \2\2 \bra{\mu}
        &&\equiv \sum_{\mu}  {c'}_{\mspace{-6mu}\mu} (\bm{\alpha}, \bm{\kappa}) \bra{\mu(\mbf{V}^*)}. \label{eq:bra_as_linear_expansion}
    \end{alignat}
\end{subequations}
The summations run over all Hartree products (or Slater determinants) within the active space.
Defining $\mbf{y} = (\bm{\alpha}, \bm{\kappa})$, we find that
\begin{align}
    \elm{\Psi'}{\partial_t}{\Psi} 
    &= \bra{\Psi'} \sum_{\mu} \big( \dot{c}_{\mu} \ket{\mu} + c_{\mu} \ket{\dot{\mu}} \big) \nn
    &= \bra{\Psi'} \sum_{\mu j}  \dot{y}_{j} \pdv{c_\mu}{y_{j}} \ket{\mu} + \bra{\Psi'} \sum_{\mu} c_{\mu} (-iG)  \ket{\mu}  \nn
    &= \sum_{j} \dot{y}_{j} \bigbraket{\Psi'}{\pdv{\Psi}{y_{j}}} - i \elm{\Psi'}{G}{\Psi}
\end{align}
and, consequently, that
\begin{align} \label{eq:complex_lagrangian}
    \mathcal{L} 
    &= \elm{\Psi'}{(i \partial_t - H)}{\Psi} \nn
    &= i \sum_{j} \dot{y}_{j} \bigbraket{\Psi'}{\pdv{\Psi}{y_{j}}} - \elm{\Psi'}{(H - G)}{\Psi} \nn
    &\equiv \mathcal{I} - \mathcal{H}'.
\end{align}
The first term depends only on $\mbf{y}$ and $\dot{\mbf{y}}$, while the second term depends
on $\mbf{y}$, $\mathbf{V}$, $\dot{\mathbf{V}}$ and $\mbf{V}^*$. 

\subsubsection{Basis set coefficients}

Equation~\eqref{eq:complex_lagrangian} has exactly the same form as the Lagrangian in Ref.~\citenum{hojlundTimedependentCoupledCluster2024a}, so we simply
restate the result from that paper: The \acp{ele}
\begin{align}
    0 &= \pdv{\bar{\mathcal{L}}}{\vplain{m}{\alpha}{q}}
    - \dv{t} \pdv{\bar{\mathcal{L}}}{\vplaindot{m}{\alpha}{q}} 
\end{align}
lead to the following equations for ${\mbf{G}^m_{\subS\subA}}$:
\begin{subequations}
    \begin{align}
        {\mbf{G}^m_{\subS\subA}} \; \mathbb{H}[\bm{\raisedrho}^m_{\!\subA\subA}] 
        &= \tfrac{1}{2} ( {\FFtildeplain{m}_{\!\subS\subA}} + {\FFtildeprimedagger{m}_{\!\!\subA\subS}}) \\
        &= \tfrac{1}{2} \mbf{V}^{m\dagger}_{\!\subS} ( \FFcheckplain{m}_{\!\!\subA} + \FFcheckprimedagger{m}_{\!\!\subA}). \label{eq:Gt_equation_halftrans}
    \end{align}
\end{subequations}
Here, $\mathbb{H}[\,\cdot\,]$ denotes the Hermitian part of a square matrix.
We have defined one-particle density matrices,
\begin{align} \label{eq:one_mode_dens_def}
    \rrho{m}{p}{q} 
    = \elm{\Psi'}{ \Etilde{m}{q}{p} }{\Psi}
    = {\sum_{\bar{p} \bar{q}}}^{(m)} \exp(\mathbf{K}^m)_{p \bar{p}} \, \elm{\psi'}{ \Etilde{m}{\bar{q}}{\bar{p}} }{\psi} \exp(-\mathbf{K}^m)_{\bar{q} q},
\end{align}
generalized mean-field or Fock matrices,
\begin{subequations} \label{eq:F_tilde_elements}
    \begin{align}
        \Ftildeprime{m}{q}{p} 
        &= \elm{\Psi'}{[H, \creatilde{m}{p}] \annitilde{m}{q}}{\Psi}, \label{eq:F_tilde_prime_elements} \\
       \Ftildeplain{m}{q}{p} 
        &= \elm{\Psi'}{\creatilde{m}{p} [\annitilde{m}{q}, H]}{\Psi}, \label{eq:F_tilde_plain_elements}
    \end{align}
\end{subequations}
and partially transformed mean-field matrices,
\begin{subequations} \label{eq:F_check_elements}
    \begin{alignat}{2}
        \Fcheckprime{m}{q}{\alpha} 
         &= \pdv{\mathcal{H}}{\vplain{m}{\alpha}{q}} 
        &&= \elm{\Psi'}{[H, \crea{m}{\alpha}] \annitilde{m}{q}}{\Psi}, \\
        \Fcheckplain{m}{\alpha}{q} 
         &= \pdv{\mathcal{H}}{\vplainconj{m}{\alpha}{q}} 
        &&= \elm{\Psi'}{\creatilde{m}{q} [\anni{m}{\alpha}, H]}{\Psi}.
    \end{alignat}
\end{subequations}
These matrices are partially transformed in the sense that one index refers to the primitive basis, while
the other refers to the orthogonal tilde basis.
Equation~\eqref{eq:one_mode_dens_def} relates the density matrix of the $\hat{\kappa}$-transformed states
to that of the untransformed states in a simple way, which is useful for practical implementations.
One also needs to address the $\hat{\kappa}$-transformation inside the mean-field
matrix elements, and some care is needed in the partially transformed case
since it involves creators and annihilators in different bases. 
It is always possible to write the (partially transformed) mean fields in
terms of contractions of (partially transformed) Hamiltonian integrals and density matrices.
The $\hat{\kappa}$-transformation adds extra one-index contractions that can be absorbed
into either the integrals or the density matrices, depending on implementation details. 
Our approach is outlined in Appendix~\ref{appendix:half_transformed_mean_fields}.

Unlike Ref.~\citenum{hojlundTimedependentCoupledCluster2024a}, we dot not solve any equations for $\mbf{G}^m_{\subA\subA}$, since
these constraint elements are simply not part of our parameterization.
Dropping the mode index for clarity,
Eq.~\eqref{eq:V_dot_general} now reads
\begin{align}
    i
    \left[
	\begin{array}{c | c}
        \dot{\mbf{V}}_{\!\!\subA} {\,} & {\,} \dot{\mbf{V}}_{\!\subS} \Tstrut
	\end{array} \right]
    =
    \left[
	\begin{array}{c | c}
        \mbf{V}_{\!\!\subA} {\,} & {\,} \mbf{V}_{\!\subS} \Tstrut
	\end{array} \right]
    \left[
	\begin{array}{c | c}
	{\mbf{0}} {\,} & {\,} {\mbf{G}_{\subA\subS}} \Tstrut
	\\ \hline
	{\mbf{G}_{\subS\subA}} {\,} & {\,} {\mbf{0}}   \Tstrut
	\end{array} \right],
\end{align}
or, focusing on the active block,
\begin{align} \label{eq:VA_dot}
    i \dot{\mbf{V}}_{\!\!\subA}
    &= \mbf{V}_{\!\subS} \, {\mbf{G}_{\subS\subA}} \nn
    &= \mbf{V}_{\!\subS}^{\phantom{\dagger}} \2 \mbf{V}_{\!\subS}^{\dagger}  \tfrac{1}{2} ( \check{\mbf{F}}_{\!\!\subA} + \check{\mbf{F}}^{\prime \dagger}_{\!\!\subA}) \, \mathbb{H}[\bm{\raisedrho}_{\!\subA\subA}]^{-1} \nn
    &= \tfrac{1}{2} \mbf{Q} ( \check{\mbf{F}}_{\!\!\subA} + \check{\mbf{F}}^{\prime \dagger}_{\!\!\subA}) \, \mathbb{H}[\bm{\raisedrho}_{\!\subA\subA}]^{-1}
\end{align}
The secondary-space projector $\mbf{Q}$ is given by
\begin{align} \label{eq:Q_def}
    \mbf{Q} = \mbf{V}_{\!\subS}^{\phantom{\dagger}} \2 \mbf{V}_{\!\subS}^{\dagger} = \mbf{1} - \mbf{V}_{\!\!\subA}^{\phantom{\dagger}} \2 \mbf{V}_{\!\!\subA}^{\dagger}
\end{align}
and allows us to propagate $\mbf{V}_{\!\!\subA}$ without reference to secondary-space quantities.

\subsubsection{\texorpdfstring{$\bm{\alpha}$}{TEXT} and \texorpdfstring{$\bm{\kappa}$}{TEXT} parameters}

The complex Lagrangian $\mathcal{L}$ is an analytic function of 
$\mbf{y}$ and $\dot{\mbf{y}}$ 
($\mbf{y}^*$ and $\dot{\mbf{y}}^*$ never appear),
so the \acp{ele}
\begin{align}
    0 &= \pdv{\bar{\mathcal{L}}}{y_i} - \dv{t} \pdv{\bar{\mathcal{L}}}{\dot{y}_i}
\end{align}
are equivalent to
\begin{align} \label{eq:eles_for_y}
    0 &= \pdv{\mathcal{L}}{y_i} - \dv{t} \pdv{\mathcal{L}}{\dot{y}_i}.
\end{align}
This set of equations is identical to that studied in Ref.~\citenum{hojlundGeneralExponentialBasis2023a},
except for the fact that the Hamiltonian appearing in $\mathcal{L}$
is modified by $G$; see Eq.~\eqref{eq:complex_lagrangian}. The operator $G$ does not depend on $\mbf{y}$ or $\dot{\mbf{y}}$,
so this difference has no consequence for the variational calculation. 
In fact, the $G$ term drops altogether due to Eq.~\eqref{eq:G_structure} and the usual killer conditions. We will thus proceed with $\mathcal{H}' \rightarrow \mathcal{H}$.
It is not difficult to show\cite{hojlundGeneralExponentialBasis2023a}
that Eq.~\eqref{eq:eles_for_y} is equivalent to
\begin{align} \label{eq:eom_simple}
    i \sum_{j} \mathcal{M}_{ij} \dot{y}_j &= \pdv{\mathcal{H}}{y_i},
\end{align}
where
\begin{equation} \label{eq:full_M_matrix_def}
    \mathcal{M}_{ij} = 
    \bigbraket{\pdv{\Psi'}{y_i}}{\pdv{\Psi}{y_j}} -
    \bigbraket{\pdv{\Psi'}{y_j}}{\pdv{\Psi}{y_i}}. 
\end{equation}
Further derivations lead to
\begin{subequations} \label{eq:eom_inverted_alpha_g_tilde}
    \begin{align}
        i \dot{\bm{\alpha}} &= \mbf{M}^{-1} \mbf{h}', \label{eq:eom_inverted_alpha} \\
        \big(\mbf{C}
        +   \mbf{A}^{\trans} \mbf{M}^{-1} \mbf{A} \big) \tilde{\mbf{g}}
        &= 
        \big( \mbf{f} + \mbf{A}^{\trans} \mbf{M}^{-1} \mbf{h} \big). \label{eq:eom_inverted_g_tilde}
    \end{align}
\end{subequations}
The matrices and vectors are defined according to
\begin{subequations} \label{eq:A_C_f_def_without_tilde}
    \begin{align} 
        M_{ij} &= 
        \bigbraket{\pdv{\psi'}{\alpha_i}}{\pdv{\psi}{\alpha_j}} -
        \bigbraket{\pdv{\psi'}{\alpha_j}}{\pdv{\psi}{\alpha_i}} \label{eq:M_matrix_def} \\
        \Aprime{i}{m}{v}{w} 
        &= \pdv{}{\alpha_i} \elm{\psi'}{\Etildeprime{m}{v}{w}}{\psi}, \label{eq:A_simple_def} \\
        \Cplain{m}{t}{u}{m}{v}{w} 
        &= \elm{\psi'}{[\Etildeprime{m}{v}{w}, \Etilde{m}{t}{u}]}{\psi}, \label{eq:C_plain_def} \\
        \fplain{m}{t}{u} &= 
        \elm{\psi'}{[\bar{H}, \Etilde{m}{t}{u}]}{\psi}, \\
        h_i
        &= \pdv{}{\alpha_i} \elm{\psi'}{\bar{H}}{\psi}, \\
        h_i'
        &= \pdv{}{\alpha_i} \elm{\psi'}{(\bar{H} - \tilde{G})}{\psi}, \\
        \bar{H} 
        &= e^{-\hat{\kappa}} H e^{\hat{\kappa}}.
    \end{align}
\end{subequations}
The operator $\tilde{G}$, which is determined by Eq.~\eqref{eq:eom_inverted_g_tilde},
is given by
\begin{align} \label{eq:G_from_kappa_dot}
    \tilde{G} 
    &= \sum_{m} {\sum_{vw}}^{(m)} \tilde{g}^{m}_{vw} \, \Etilde{m}{v}{w} \nn
    &= i e^{-\hat{\kappa}} \dv{e^{\hat{\kappa}}}{t} \nn
    &= i \sum_{m} {\sum_{tu}}^{(m)} \kappaplaindot{m}{t}{u} \, e^{-\hat{\kappa}} \pdv{e^{\hat{\kappa}}}{\kappaplain{m}{t}{u}} \nn
    &\equiv i \sum_{m} {\sum_{tu}}^{(m)} \kappaplaindot{m}{t}{u} \, D^{m}_{tu} \nn
    &= i \sum_{m} {\sum_{tuvw}}^{(m)} \kappaplaindot{m}{t}{u} \, \dplain{m}{t}{u}{v}{w} \Etilde{m}{v}{w},
\end{align}
or, in matrix-vector notation,
\begin{align}
    \tilde{\mbf{g}}^m = i (\mbf{D}^m)^{\trans} \dot{\bm{\kappa}}^m_{\subA\subA}.
\end{align}
Reference~\citenum{hojlundGeneralExponentialBasis2023a} shows how this
equation can be solved for $\dot{\bm{\kappa}}^m_{\subA\subA}$ without explicitly constructing and inverting the
large ($N_{\!\subA}^2 \times N_{\!\subA}^2$) matrix $\mbf{D}^m$. We note that it is also perfectly possible to
propagate the matrices
\begin{subequations}
    \begin{align}
        \mbf{U}^m_{\!\subA\subA}   &= \exp(\mbf{K}^m_{\subA\subA}), \\ 
        \mbf{W}^m_{\!\!\subA\subA} &= \exp(-\mbf{K}^m_{\subA\subA}),
    \end{align}
\end{subequations}
rather than $\mbf{K}^m$ itself. The second line of Eq.~\eqref{eq:G_from_kappa_dot} implies
\begin{align}
    \tilde{\mbf{G}}^m = i \mbf{W}^m_{\!\!\subA\subA} \dot{\mbf{U}}^m_{\!\subA\subA} = - i \dot{\mbf{W}}^m_{\!\!\subA\subA} \mbf{U}^m_{\!\subA\subA} 
\end{align}
or, using the fact that $\mbf{W}^m$ and $\mbf{U}^m$ are each other's inverses,
\begin{subequations}
    \begin{align}
        \dot{\mbf{U}}^m_{\!\subA\subA}   &= -i \mbf{U}^m_{\!\subA\subA} \tilde{\mbf{G}}^m, \\ 
        \dot{\mbf{W}}^m_{\!\!\subA\subA} &= +i \tilde{\mbf{G}}^m \mbf{W}^m_{\!\!\subA\subA}.
    \end{align}
\end{subequations}
The exponential and linear formulations are of course fully equivalent and require the
solution of the same set of linear equations, Eq.~\eqref{eq:eom_inverted_g_tilde}. 
\subsection{Working equations for the CC \textit{Ansatz}} \label{sec:working_equations}
We now state the working equations in greater detail with emphasis on the \ac{cc} \textit{Ansatz}:
\begin{subequations} \label{eq:cc_ansatz}
\begin{gather}
    \ket{\Psi}  = e^{\hat{\kappa}} \ket{\psi}  = e^{\hat{\kappa}} e^{T} \ket{\Phi}, \\
    \bra{\Psi'} = \bra{\psi'} e^{-\hat{\kappa}} = \bra{\Phi} L e^{-T} e^{-\hat{\kappa}}.
\end{gather}
\end{subequations}
The operators $T$ and $L$ are given by
\begin{subequations}
    \begin{alignat}{2}
        T &= \sum_{\mu} t_{\mu} \tilde{\tau}_{\mu}           &&= t_0 + T_2 + T_3 + \ldots \\
        L &= \sum_{\mu} l_{\mu} \tilde{\tau}_{\mu}^{\dagger} &&= l_0 + L_2 + L_3 + \ldots
    \end{alignat}
\end{subequations}
where it should be noted that the single excitations are omitted since they are redundant
with the basis set transformation generated by $\hat{\kappa}$.\cite{kvaalInitioQuantumDynamics2012,satoCommunicationTimedependentOptimized2018,madsenTimedependentVibrationalCoupled2020,pedersenGaugeInvariantCoupled1999,pedersenGaugeInvariantCoupled2001}
The excitation operators $\tilde{\tau}_{\mu}$ and the reference $\ket{\Phi}$
are defined in terms of the moving creators $\creatilde{m}{v}$,
and we can thus consider Eqs.~\eqref{eq:cc_ansatz} as an \ac{nocc}-type expansion inside an evolving active space.
This \textit{Ansatz} is able to reproduce the complete wave function expansion
within the active space,\cite{myhreDemonstratingThatNonorthogonal2018}
which means that exact active-space densities and mean fields are obtained when $T$ and $L$ are not truncated.
Exact densities and mean fields in turn generate the exact evolution of the active space itself
according to Eq.~\eqref{eq:VA_dot}. The \textit{Ansatz} thus reproduces the \ac{mctdh} limit in the
vibrational case and the \ac{mctdhf} limit in the electronic structure case.

The amplitude \acp{eom} can of course be derived by applying the \ac{tdbvp}
directly to the \ac{cc} \textit{Ansatz}, but in our general framework we order the
amplitudes like $\bm{\alpha} = (\mbf{t}, \mbf{l})$ and apply Eq.~\eqref{eq:M_matrix_def} to find that
\begin{align} \label{eq:M_and_M_inv}
    \mbf{M} = 
    \left[
	\begin{array}{c | c} 
	 \mbf{0} & -\mbf{1} \\ \hline
	+\mbf{1} &  \mbf{0} 
	\end{array} \right], \quad
    \mbf{M}^{-1} = 
    \left[
	\begin{array}{c | c} 
	 \mbf{0} & +\mbf{1} \\ \hline
	-\mbf{1} &  \mbf{0} 
	\end{array} \right].
\end{align}
Having obtained $\mbf{M}^{-1}$, Eq.~\eqref{eq:eom_inverted_alpha} readily yields the result:
\begin{subequations}
    \begin{align}
        i \dot{t}_{\mu} &= + \elm{\mu}{e^{-T}(\bar{H} - \tilde{G})}{\psi}, \\
        i \dot{l}_{\mu} &= - \elm{\psi'}{[\bar{H} - \tilde{G}, \tilde{\tau}_{\mu}]}{\psi}.
    \end{align}
\end{subequations}
The non-unitary basis set transformation is determined by Eq.~\eqref{eq:eom_inverted_g_tilde},
\begin{align}
    \big(\mbf{C} + \mbf{A}^{\trans} \mbf{M}^{-1} \mbf{A} \big) \tilde{\mbf{g}}
    =  \big( \mbf{f} + \mbf{A}^{\trans} \mbf{M}^{-1} \mbf{h} \big),
\end{align}
which we rewrite simply as
\begin{align}
    \mbf{C}' \tilde{\mbf{g}} = \mbf{f}'.
\end{align}
The primes indicate the inclusion of the terms depending on $\mathbf{A}$.
Referring to Eqs.~\eqref{eq:A_C_f_def_without_tilde}, one easily computes
the elements of $\mbf{C}'$ and $\mbf{f}'$:
\begin{multline} \label{eq:tdmvcc_active_space_c}
    \Cprime{m}{t}{u}{m}{v}{w} =
    \elm{\psi'}{[\Etildeprime{m}{v}{w}, \Etilde{m}{t}{u}]}{\psi} \\
        +\sum_\mu \Big( 
        \elm{\psi'}{[\Etilde{m}{t}{u}, \tilde{\tau}_\mu]}{\psi}   \elm{\mu}{e^{-T} \Etildeprime{m}{v}{w}}{\psi} 
        - \elm{\psi'}{[\Etildeprime{m}{v}{w}, \tilde{\tau}_\mu]}{\psi}   \elm{\mu}{e^{-T} \Etilde{m}{t}{u}}{\psi}
        \Big)
\end{multline}
and
\begin{multline} \label{eq:tdmvcc_active_space_f}
    \fprime{m}{t}{u} = 
    \elm{\psi'}{[\bar{H}, \Etilde{m}{t}{u}]}{\psi} \\
        + \sum_\mu \Big( 
        \elm{\psi'}{[\Etilde{m}{t}{u}, \tilde{\tau}_\mu]}{\psi}   \elm{\mu}{e^{-T} \bar{H}}{\psi} 
        - \elm{\psi'}{[\bar{H}, \tilde{\tau}_\mu]}{\psi}   \elm{\mu}{e^{-T} \Etilde{m}{t}{u}}{\psi}
        \Big).
\end{multline}
The active space is divided into occupied ($i,j,\ldots$) and virtual ($a,b,\ldots$) subspaces, 
which means that index pairs $tu$ can be divided into four types: 
$ai$ (up), $ia$ (down), $ab$ (forward) and $ij$ (passive). A particular feature
of vibrational structure theory is that each mode has only a single occupied index,
something that is useful in concrete implementations. Using $u$, $d$, $f$ and $p$
to label the four types of index pairs induces a block structure in the linear equations.
Many of these blocks can be shown to vanish,\cite{madsenTimedependentVibrationalCoupled2020}
so that we are left with
\begin{align} \label{eq:biorthogonal_tdmvcc_constraint_equations_structure}
	\left[
	\begin{array}{c | c | c | c} 
    \mbf{0}          & {^{ud}\mbf{C}}  & \mbf{0} & \mbf{0} \TTstrut
	\\ \hline
	{^{du}\mbf{C}}   & {^{dd}\mbf{C}}' & \mbf{0} & \mbf{0} \TTstrut
	\\ \hline
	\mbf{0}     	                & \mbf{0} & \mbf{0} & \mbf{0} \TTstrut
	\\ \hline
	\mbf{0}     	                & \mbf{0} & {\hspace{1.7ex}}\mbf{0}{\hspace{1.7ex}} & \hspace{1.7ex}\mbf{0}{\hspace{1.5ex}} \TTstrut
	\end{array}\right] 
    \left[
    \begin{array}{c} 
    {^{u}\tilde{\mbf{g}}} \TTstrut
    \\ \hline
    {^{d}\tilde{\mbf{g}}} \TTstrut
    \\ \hline
    {^{f\!}\tilde{\mbf{g}}} \TTstrut
    \\ \hline
    {^{p}\tilde{\mbf{g}}} \TTstrut
    \end{array}\right] 
	=
	\left[
	\begin{array}{c} 
	{^u\mbf{f} } \TTstrut
	\\ \hline
	{^d\mbf{f}'} \TTstrut
	\\ \hline
	\mbf{0} \TTstrut
	\\ \hline
	\mbf{0} \TTstrut
	\end{array}\right].
\end{align}
Evidently, the forward and passive blocks of $\tilde{\mbf{g}}$ are redundant. This is not
surprising since these blocks generate transformations within the virtual and occupied
spaces, respectively. Such intraspace transformations are redundant for a wide
range of time-dependent and time-independent wave function \textit{Ansätze}, including \ac{cc}. 
One can in fact determine the redundancies from the outset by geometric arguments
(see e.g. Refs.\citenum{kvaalInitioQuantumDynamics2012} and \citenum{miyagiTimedependentRestrictedactivespaceSelfconsistentfield2013}), 
but we find it satisfying and reassuring to see that they also show up in a concrete derivation.

The non-redundant equations now read
\begin{subequations}
    \begin{gather}
        {^{ud}\mbf{C}} \,  {^{d}\tilde{\mbf{g}}} = {^u\mbf{f} }, \\
        {^{du}\mbf{C}} \,  {^u\tilde{\mbf{g}} } + {^{dd}\mbf{C}}'  \, {^{d}\tilde{\mbf{g}}} = {^d\mbf{f}'},
    \end{gather}
\end{subequations}
and can be solved as
\begin{subequations} \label{eq:gd_gu_equations}
\begin{gather}
    {^{d}\mbf{g}} = ({^{ud}\mbf{C}})^{-1} \, {^u\mbf{f} }, \\
    {^u\mbf{g} } = ({^{du}\mbf{C}})^{-1} ( {^d\mbf{f}'} - {^{dd}\mbf{C}}'  \, {^{d}\mbf{g}} ).
\end{gather}
\end{subequations}
We observe the following: (i) the \textit{up} and \textit{down} equations
can be solved sequentially and (ii) the matrices ${^{ud}\mbf{C}}$ and ${^{du}\mbf{C}}$ 
are diagonal in the mode index [see Eq.~\eqref{eq:C_plain_def}], meaning they can be
inverted one mode at a time. Equations~\eqref{eq:gd_gu_equations}
thus consist of $2M$ linear systems of dimension $N_{\subV} N_{\subO} \times N_{\subV} N_{\subO} = N_{\subV} \times N_{\subV}$,
where $M$ is the number of vibrational modes. The symbols $N_{\subV}$ and $N_{\subO} = 1$ denote the 
number of virtual and occupied functions per mode, and we assume, for simplicity, 
that $N_{\subV}$ is the same for all modes (this is often not the case in practical calculations).
The fact that the linear equations are up-down and mode-mode
decoupled constitutes a large reduction in the computational effort compared
to the orthogonal case,\cite{hojlundTimedependentCoupledCluster2024a} 
where one generally (except at the doubles level) has to solve a single linear system of dimension 
$2 M N_{\subV} N_{\subO} \times 2 M N_{\subV} N_{\subO} = 2 M N_{\subV} \times 2 M N_{\subV}$.
In electronic structure theory, only the up-down decoupling applies, and
the reduction from one system of dimension $2 N_{\subV} N_{\subO} \times 2 N_{\subV} N_{\subO}$ (orthogonal orbitals) to 
two systems of dimension $N_{\subV} N_{\subO} \times N_{\subV} N_{\subO}$ (biorthogonal orbitals) is less striking, although still significant.


 

\section{IMPLEMENTATION} \label{sec:implementation}
The sTDMVCC method has been implemented in the \ac{midas},\cite{MidasCpp2023040}
which includes an array of methods for solving the time-independent and
time-dependent vibrational Schrödinger equations and for constructing \acp{pes}.
At the two-mode coupling level (sTDMVCC$[2]$/H2, i.e. $T = T_2$ and $H = H_1 + H_2$),
we have simply adapted the efficient TDMVCC$[2]$/H2 implementation of 
Ref.~\citenum{jensenEfficientTimedependentVibrational2023a}, which
offers a cubic-scaling computational effort. For general coupling levels in $T$ and
$H$ we adapt the original \ac{tdmvcc} code,\cite{madsenTimedependentVibrationalCoupled2020}
which is based on the \ac{fsmr} implementation of Ref.~\citenum{hansenExtendedVibrationalCoupled2020}.
This is a FCI-type code, so the computational cost scales exponentially with respect to system size.
Such a scaling is of course prohibitive for practical usage, but the \ac{fsmr} code
has nevertheless been instrumental for testing and benchmarking new \ac{cc} variants
with limited programming effort.

To avoid possible (near-)singularities in the one-mode density matrices [see Eq.\eqref{eq:VA_dot}],
we use a regularization scheme based on \ac{svd}:
\begin{align}
    \bm{\raisedrho}_{\!\subA\subA} = \mathbf{X} \mathbf{\Sigma} \mathbf{Y}^\dagger
    \rightarrow \mathbf{X} \big( \mathbf{\Sigma} + \epsilon \exp(-\mathbf{\Sigma}/\epsilon)  \big) \mathbf{Y}^\dagger.
\end{align}
Here, $\epsilon$ is a small regularization parameter (typically, $\epsilon = 10^{-8}$). 
We rarely observe singularities in the main body of a simulation, 
but they do occur close to $t = 0$ when using a \ac{vscf} initial state or some other initial
state that has active basis functions with exactly zero occupation.
The same kind of regularization is used for inverting the individual blocks of
${^{ud}\mathbf{C}}$ and ${^{du}\mathbf{C}}$ in Eqs.~\eqref{eq:gd_gu_equations}.
Finally, we modify the secondary-space projector of Eq.~\eqref{eq:Q_def}
according to
\begin{align}
    \mbf{Q}   = \mbf{1} - \mbf{V}_{\!\!\subA}^{\phantom{\dagger}} \2 \mbf{V}_{\!\!\subA}^{\dagger}
    \rightarrow \mbf{1} - \mbf{V}_{\!\!\subA}^{\phantom{\dagger}} (\mbf{V}_{\!\!\subA}^{\dagger} \mbf{V}_{\!\!\subA}^{\phantom{\dagger}})^{-1} \mbf{V}_{\!\!\subA}^{\dagger}.
\end{align}
This guarantees that $\mbf{Q}$ remains a proper projector even if the columns of $\mbf{V}_{\!\!\subA}$ are not strictly orthonormal.
 
\section{NUMERICAL EXAMPLES} \label{sec:results}
We consider three examples in order to study the numerical stability and convergence properties of
the sTDMVCC method: 
(i) \Ac{ivr} in water, 
(ii) the intramolecular proton transfer in a 6D salicylaldimine model,\cite{polyakCompleteDescriptionTunnelling2015} and
(iii) the dynamics after an $S_1 \rightarrow S_0$ transition in a 5D \textit{trans}-bithiophene model.\cite{madsenVibrationallyResolvedEmission2019} 
We employ basis set division (i.e., $N_{\!\subA} < N$) in all cases and compare with \ac{mctdh} reference
calculations as well as TDMVCC, rpTDMVCC, and oTDMVCC calculations. The \acp{eom} are
integrated using the \ac{dop853}\cite{hairerSolvingOrdinaryDifferential2009b} algorithm
with tight absolute and relative tolerances for controlling the step size ($\tau_{\mathrm{abs}} = \tau_{\mathrm{rel}} = 10^{-12}$).
Density matrices are regularized using $\epsilon = 10^{-8}$.

We take the physical expectation value of a Hermitian operator $\Omega$ to be
\begin{align}
    \langle \Omega \rangle = \mathrm{Re} \elm{\Psi'}{\Omega}{\Psi}.
\end{align}
The corresponding imaginary part has no physical or experimental meaning as such, but
it can be used to assess how well the wave function agrees with the exact limit
where the imaginary part vanishes identically. We will thus write 
$\langle \Omega \rangle_{\mathrm{Re}} = \mathrm{Re} \elm{\Psi'}{\Omega}{\Psi}$ and 
$\langle \Omega \rangle_{\mathrm{Im}} = \mathrm{Im} \elm{\Psi'}{\Omega}{\Psi}$
when comparing the real and imaginary parts.

We also report \acp{acf} computed as\cite{hansenTimedependentVibrationalCoupled2019}
\begin{align}
    S(t) = \braket{\Psi'(0) | \Psi(t)},
\end{align}
and, occasionally, basis set non-orthogonality defined as
\begin{subequations}
\begin{gather}
    \eta_{\mathrm{ket}}^m = || (\bm{\mathcal{U}}^m_{\!\subA})^{\dagger}   (\bm{\mathcal{U}}^m_{\!\subA})              - \mathbf{1} ||, \\
    \eta_{\mathrm{bra}}^m = || (\bm{\mathcal{W}}^m_{\!\!\subA})           (\bm{\mathcal{W}}^m_{\!\!\subA})^{\dagger}  - \mathbf{1} ||,
\end{gather}
\end{subequations}
where $|| \cdot ||$ is the Frobenius norm.

\subsection{Water}

The first example is the \ac{ivr} in water after exciting the symmetric stretch to
$n = 2$. We model this process by taking the $[0,2,0]$ state on the harmonic part of the \ac{pes}
as the initial state, followed by relaxation on the anharmonic and coupled \ac{pes}.
The calculation is repeated at the MCTDH, rpTDMVCC[2--3], oTDMVCC[2--3], sTDMVCC[2--3], and TDMVCC[2--3] levels
with $N = 20$ primitive modals per mode and a range of $N_{\!\subA}$ from $2$ to $20$.
The propagation time is set to \SI{20000}{au} ($\sim$\SI{484}{fs}), but the number of integrator steps is capped
at $\num{90000}$ in order to allow the program to terminate in an orderly fashion if the
step size drops by a large amount.

\subsubsection{Numerical stability}

\begin{figure}[H]
    \centering

    \begin{subfigure}[c]{0.5\textwidth}
        \centering
        \caption{}
        \includegraphics[width=\textwidth]{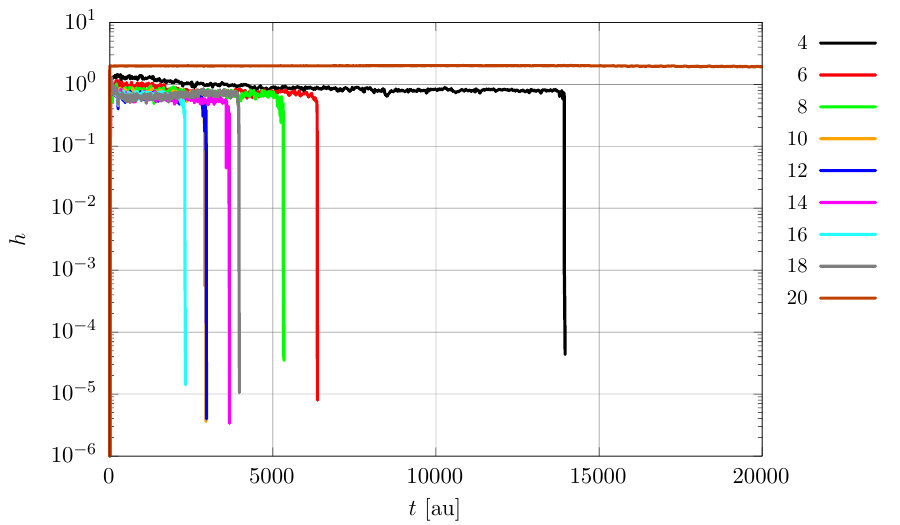}
        \label{fig:water_tdmvcc2_stepsize}
    \end{subfigure}

    \begin{subfigure}[c]{0.5\textwidth}
        \centering
        \caption{}
        \includegraphics[width=\textwidth]{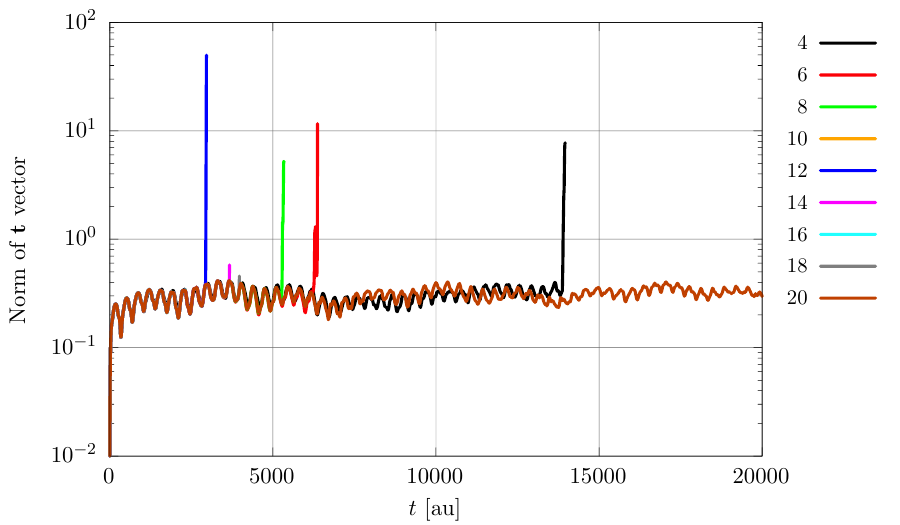}
        \label{fig:water_tdmvcc2_norm_t}
    \end{subfigure}

    \begin{subfigure}[c]{0.5\textwidth}
        \centering
        \caption{}
        \includegraphics[width=\textwidth]{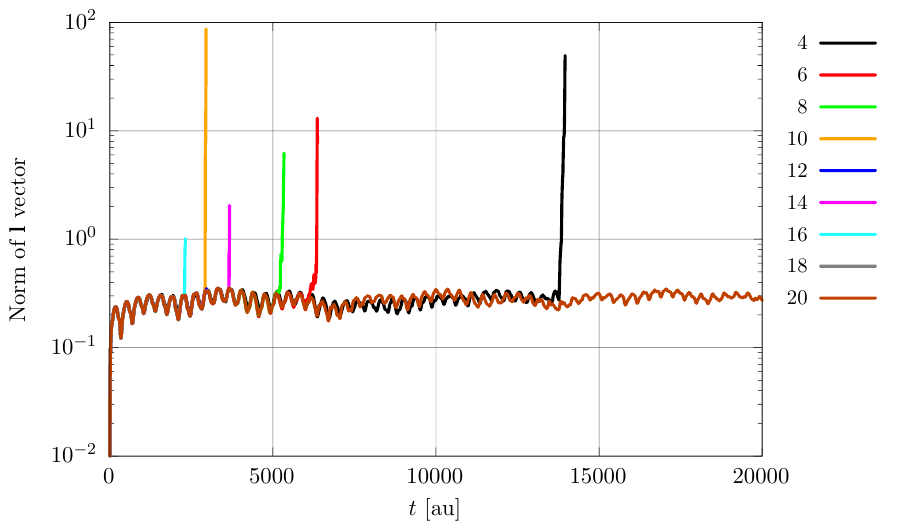}
        \label{fig:water_tdmvcc2_norm_l}
    \end{subfigure}

    \caption{\Ac{ivr} in water at the TDMVCC[2] level with $N = 20$ and a range of $N_{\!\subA}$ (as indicated by the figure legends).
    (a) Integrator step size. (b) Norm of $\mathbf{t}$ vector. (c) Norm of $\mathbf{l}$ vector. }
    \label{fig:water_tdmvcc2_stepsize_norms}
\end{figure}

\begin{figure}[H]
    \centering
    \includegraphics[width=\textwidth]{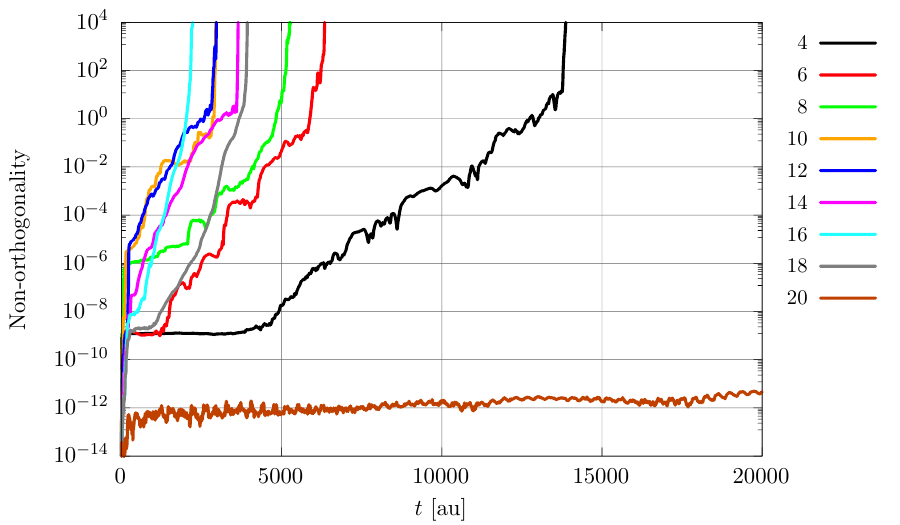}

    \caption{\Ac{ivr} in water at the TDMVCC[2] level with $N = 20$ and a range of $N_{\!\subA}$ (as indicated by the figure legends).
    Ket non-orthogonality for mode $m = 2$ (bend).}
    \label{fig:water_tdmvcc2_nonortho_ket_mode_2}
\end{figure}

All simulations proceed in an inconspicuous and stable manner, 
except at the TDMVCC[2] and TDMVCC[3] levels.
These calculations
suffer from the kind of instability that we first described in Ref.~\citenum{hojlundBivariationalTimedependentWave2022},
as illustrated by Fig.~\ref{fig:water_tdmvcc2_stepsize_norms}. Figure~\ref{fig:water_tdmvcc2_stepsize}
shows the integrator step size as a function of time for the TDMVCC[2] calculations with varying $N_{\!\subA}$.
Whenever $N_{\!\subA} < N = 20$, the step size eventually drops, making the equations virtually impossible 
to integrate. We observe, furthermore, that the sharp drop in step size correlates with a sharp increase in the amplitude vector norms
[Figs.~\ref{fig:water_tdmvcc2_norm_t} and \ref{fig:water_tdmvcc2_norm_l}].
Looking at Fig.~\ref{fig:water_tdmvcc2_stepsize_norms}, it appears that the breakdown happens suddenly and without
warning, but the ket-state non-orthogonality [Fig.~\ref{fig:water_tdmvcc2_nonortho_ket_mode_2}]
shows that this is not quite the case. Here, we note a steady build-up of
non-orthogonality followed by a violent increase that eventually leads to breakdown. 
It should be emphasized that this only happens when $N_{\!\subA} < N$,
i.e. when the active bra and ket functions are allowed to span different spaces.

Similar problematic behavior is observed for TDMVCC[3] (see Figs.~S1 and S2 in the supplementary material),
while all remaining calculations run to completion without any problems (see Figs.~S3--S14).

\subsubsection{\Aclp{acf}}

\begin{figure}[H]
    \centering
    \includegraphics[width=0.85\textwidth]{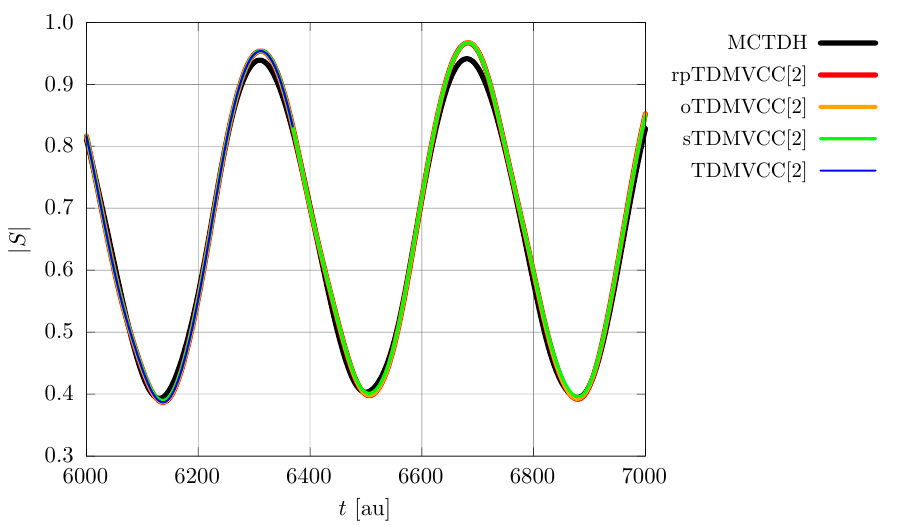}

    \caption{\Ac{ivr} in water at the MCTDH, TDMVCC[2], sTDMVCC[2], oTDMVCC[2], and rpTDMVCC[2] levels with $N = 20$ and $N_{\!\subA} = 6$.
    \Aclp{acf}. Note that the TDMVCC[2] trace terminates at $t = \SI{6369}{au}$.}
    \label{fig:water_xtdmvcc2_autocorr}
\end{figure}

One might fear that the erratic numerical behavior of the TDMVCC calculations would also
lead to an erratic description of observables. This is, however, not the case, as exemplified by
Fig.~\ref{fig:water_xtdmvcc2_autocorr}, which shows the \ac{acf} at the MCTDH, TDMVCC[2], sTDMVCC[2], oTDMVCC[2],
and rpTDMVCC[2] levels of theory with $N = 20$ and $N_{\!\subA} = 6$. All \ac{cc} variants agree closely
with each other, and they reproduce the \ac{mctdh} reference result quite well. This shows that the \ac{cc}
\textit{Ansatz} as such is well suited for describing the dynamics at hand. 
Nevertheless, the TDMVCC[2] calculation breaks down at $t = \SI{6369}{au}$, while the remaining variants remain stable.
We interpret this as a sign that the TDMVCC instability is not caused by the physics of the problem
or by shortcomings of \ac{cc} expansion in itself, but by the mathematical peculiarities of the TDMVCC basis set.

\subsection{6D salicylaldimine}

Our second example is the intra-molecular proton transfer in salicylaldimine.
This is an asymmetric double-well system with the lower well corresponding to the enol
tautomer, which is more stable than the keto tautomer.
We use the six-dimensional model \ac{pes} from Ref.~\citenum{polyakCompleteDescriptionTunnelling2015},
which includes the modes labelled as $Q_1$, $Q_{10}$, $Q_{11}$, $Q_{13}$, $Q_{32}$, and $Q_{36}$
($Q_1$ is the double-well mode).
The initial state is a Gaussian wave packet placed just on the enol side of the barrier,
but with an energy lower than the barrier height 
(see Refs.~\citenum{polyakCompleteDescriptionTunnelling2015} and \citenum{madsenMRMCTDHFlexible2020} for details).
Crossing the barrier is thus classically forbidden and any flux over the barrier is solely due to tunneling.

The primitive basis for the calculation is obtained in the following way: 
Initially, a large B-spline basis is used for the double-well mode,
while the remaining modes are described by 21 harmonic oscillator functions each.
This basis is then transformed to a set of orthonormal \ac{vscf}
modals. For the double-well mode, we use the lowest 30 \ac{vscf} modals
as the primitive basis, while the remaining modes use all 21 \ac{vscf} modals.
The initial active modals are simply the $N_{\!\subA} = 3$ lowest \ac{vscf} modals
for each mode. We do not expect the calculations to be fully converged with respect
to $N_{\!\subA}$, but we note that our results are very similar
to Ref.~\citenum{madsenMRMCTDHFlexible2020}. Even if $N_{\!\subA}$
is not quite converged, this has no effect on the comparison between methods.

The total simulation time is \SI{20000}{au} ($\sim$\SI{484}{fs}), which
is significantly longer than the \SI{4200}{au} ($\sim$\SI{100}{fs}) of Refs.~\citenum{polyakCompleteDescriptionTunnelling2015} and \citenum{madsenMRMCTDHFlexible2020}.
We have prioritized long simulations in order to assess numerical stability at long integration times. The calculation is repeated at the
MCTDH, rpTDMVCC[2--6], oTDMVCC[2--6], sTDMVCC[2--6], and TDMVCC[2--6] levels.

\subsubsection{Numerical stability}

\begin{figure}[H]
    \centering
    \includegraphics[width=0.85\textwidth]{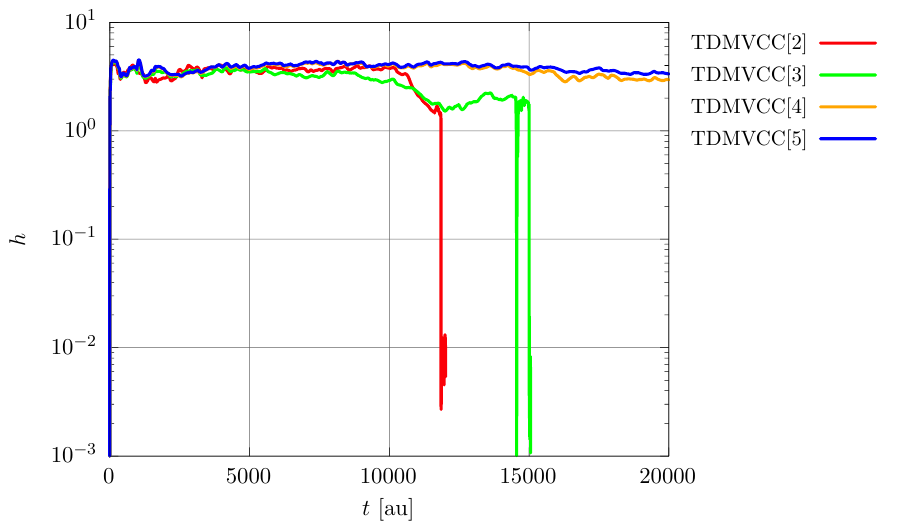}

    \caption{Integrator step size for 6D salicylaldimine
    at the TDMVCC[2--6] levels.
    The TDMVCC[2], TDMVCC[3], and TDMVCC[4] calculations terminated at $t = \SI{6540}{au}$, $t = \SI{4989}{au}$, and $t = \SI{13413}{au}$, respectively.}
    \label{fig:sal6_stepsize}
\end{figure}

Stability problems are observed for TDMVCC[2--4] as 
demonstrated by the integrator step size in Fig.~\ref{fig:sal6_stepsize}.
TDMVCC[5] and TDMVCC[6] remain stable within the time interval we have considered,
but it is quite possible that they will fail at even longer integration times.
The remaining \ac{cc} variants integrate smoothly to completion
and are not shown here (see Figs.~S15--S17).

\subsubsection{Expectation values}

The flux over the transition state (which is the zero of all coordinates)
can be quantified by the flux operator\cite{beckMulticonfigurationTimedependentHartree2000} for mode $Q_1$ at the point $Q_1 = 0$:
\begin{align}
    F_1 = i [\hat{H}, \Theta(Q_1)] = -\frac{i}{2} \left( \pdv{}{Q_1} \delta(Q_1) + \delta(Q_1) \pdv{}{Q_1}  \right).
\end{align}
Here, $\Theta(Q_1)$ is the Heaviside step function, $\delta(Q_1)$ is the Dirac delta function, and
the second expression holds only in rectilinear coordinates.\cite{madsenMRMCTDHFlexible2020}
Negative flux corresponds to the wave function crossing from the enol side to the keto side and vice versa.

\begin{figure}[H]
    \centering

    \begin{subfigure}[c]{0.69\textwidth}
        \centering
        \caption{}
        \includegraphics[width=\textwidth]{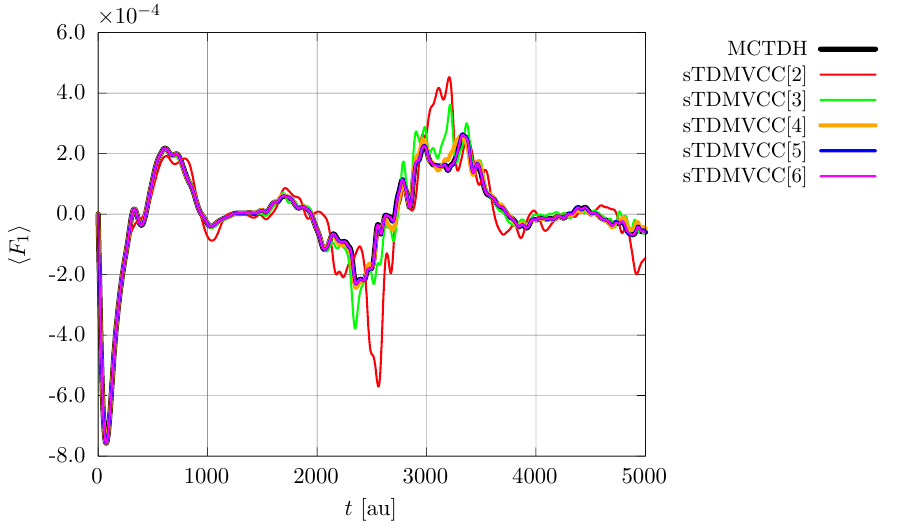}
        \label{fig:sal6_F1_short}
    \end{subfigure}

    \begin{subfigure}[c]{0.69\textwidth}
        \centering
        \caption{}
        \includegraphics[width=\textwidth]{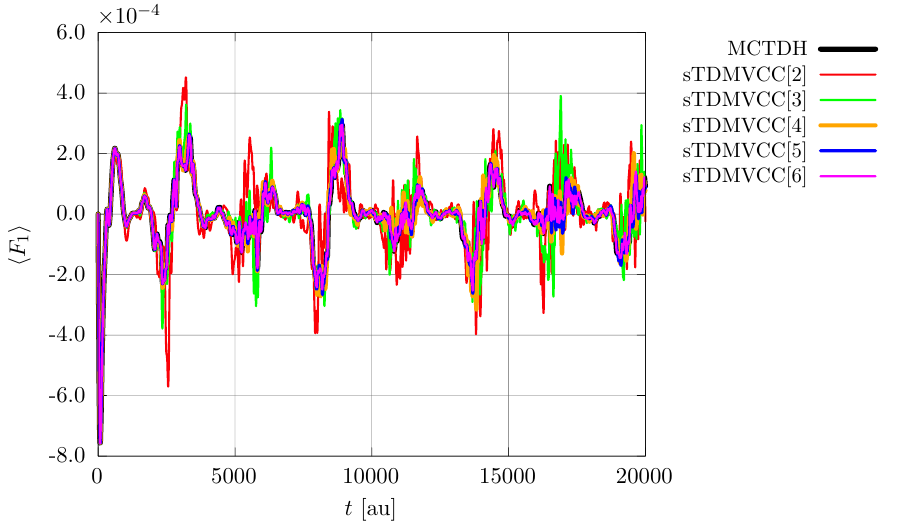}
        \label{fig:sal6_F1_long}
    \end{subfigure}

    \caption{Expectation value $\langle F_{1} \rangle$ for 6D salicylaldimine
    at the MCDTH and sTDMVCC[2--6] levels. (a) Short times. (b) Full time interval.}
    \label{fig:sal6_F1}
\end{figure}

Figure~\ref{fig:sal6_F1} shows the flux for sTDMVCC[2--6] compared to the MCTDH result.
The flux is a sensitive property, and the double-well system is rather challenging, so
we are able to see clear differences between different excitation levels.
We find the quality of the sTDMVCC results to be quite acceptable already at the doubles level,
but sTDMVCC[5] is in fact needed to get results that lie on top of MCTDH. Convergence can
most likely be improved by a flexible excitation-space scheme similar to that of Ref.~\citenum{madsenMRMCTDHFlexible2020},
where one includes, for example, all triple excitations involving the double-well mode in addition to all
double excitations.

\begin{figure}[H]
    \centering

    \begin{subfigure}[c]{0.75\textwidth}
        \centering
        \caption{Average error}
        \includegraphics[width=\textwidth]{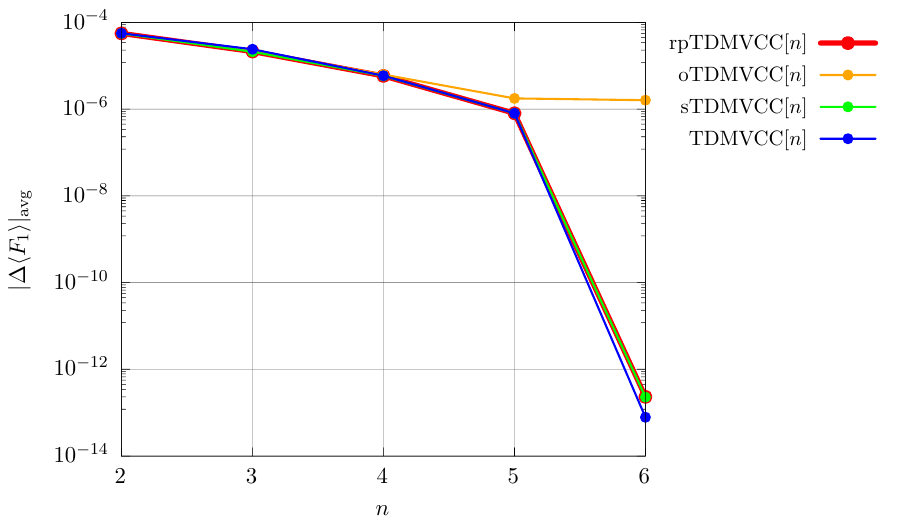}
        \label{fig:sal6_F1_diff}
    \end{subfigure}

    \begin{subfigure}[c]{0.75\textwidth}
        \centering
        \caption{Average imaginary vs. real part}
        \includegraphics[width=\textwidth]{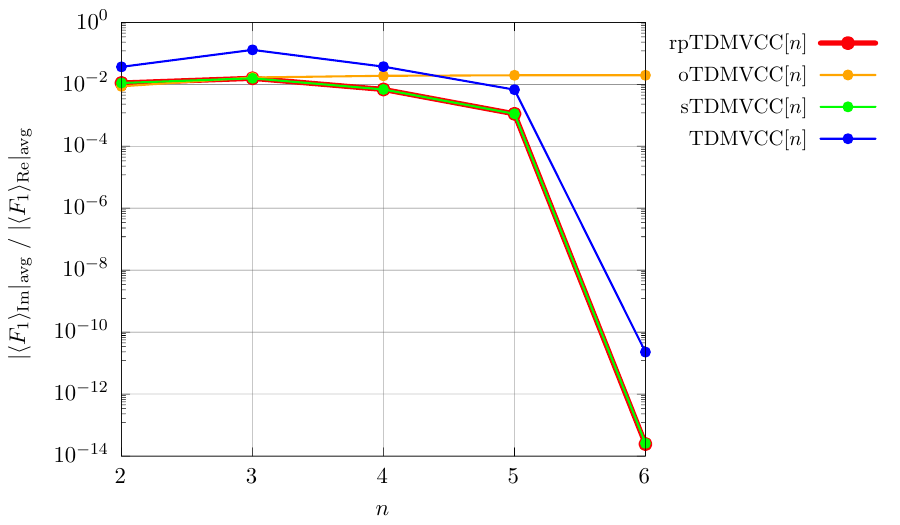}
        \label{fig:sal6_F1_re_im}
    \end{subfigure}

    \caption{Expectation value $\langle F_{1} \rangle$ for 6D salicylaldimine
    at the rpTDMVCC[2--6], sTDMVCC[2--6], oTDMVCC[2--6] and TDMVCC[2--6] levels.
    (a) Average error relative to MCTDH. (b) Average imaginary part divided by average real part.
    The averaging has been done over the time interval from $t = \SI{50}{au}$ to $t = \SI{4989}{au}$.}
    \label{fig:sal6_F1_diff_re_im}
\end{figure}

To the unaided eye, the other \ac{cc} variants appear similar to sTDMVCC (see Figs.~S18--S20),
so we turn to the average absolute error [Fig.~\ref{fig:sal6_F1_diff}] for a definite ranking.
The \acp{eom} are initially somewhat
difficult to integrate due to the singular density matrices of the \ac{vscf} initial state,
resulting in oscillations of the absolute error at short times. We thus exclude the first \SI{50}{au}
from the averaging. The averaging is terminated at $t = \SI{4989}{au}$ (where TDMVCC[3] fails)
in order to obtain an unbiased comparison.
We observe that oTDMVCC does not converge to MCTDH, although this deficiency only shows up
at the quintuple and sextuple levels. The remaining variants do converge, and the convergence
patterns are very similar. In practical calculations on
salicylaldimine or similar systems, one would thus not observe significant
differences in quality between the methods.

As an additional check on the \ac{cc} wave functions, we report the average value of $|\langle F_1 \rangle_{\mathrm{Im}}|$
divided by the average value of $|\langle F_1 \rangle_{\mathrm{Re}}|$ [Fig.~\ref{fig:sal6_F1_re_im}].
This provides a simple measure of the relative magnitude of the unphysical imaginary part.
For the rpTDMVCC, sTDMVCC, and TDMVCC methods, which all allow non-unitary basis set transformations
within the active space, the imaginary part goes to zero when the cluster expansion becomes complete.
This is not the case for the oTDMVCC method, which is another clear sign that the orthogonal hierarchy cannot
reproduce the exact solution correctly. 
Expectation values of the displacement coordinates behave similarly (see Figs.~S21--S50).

It is also interesting to note that the unphysical imaginary part decreases quite slowly as a function of excitation level (rpTDMVCC and sTDMVCC) or not
at all (oTDMVCC). This pattern differs qualitatively from the error in $\langle F_1 \rangle$ [Fig.~\ref{fig:sal6_F1_diff}], which
improves quite steadily with respect to excitation level.

\subsubsection{Energy conservation}

\begin{figure}[H]
    \centering

    \begin{subfigure}[c]{0.75\textwidth}
        \centering
        \caption{Real part}
        \includegraphics[width=\textwidth]{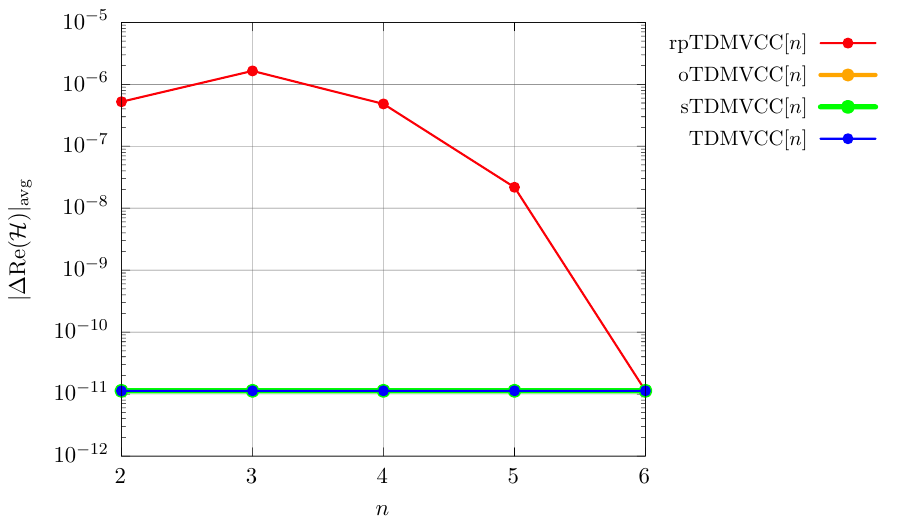}
        \label{fig:sal_energy_re}
    \end{subfigure}

    \begin{subfigure}[c]{0.75\textwidth}
        \centering
        \caption{Imaginary part}
        \includegraphics[width=\textwidth]{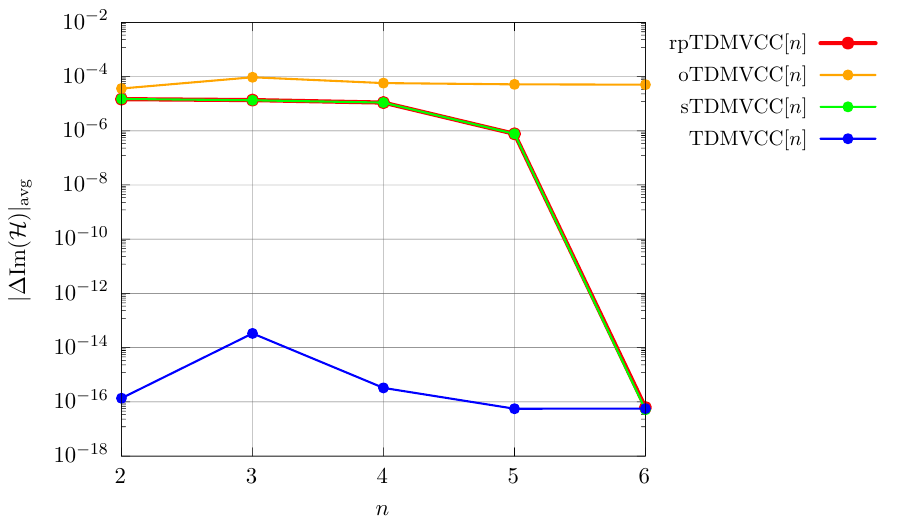}
        \label{fig:sal_energy_im}
    \end{subfigure}

    \caption{Energy conservation for 6D salicylaldimine
    at the rpTDMVCC[2--6], sTDMVCC[2--6], oTDMVCC[2--6] and TDMVCC[2--6] levels. 
    (a) Average real part (this is the physical energy).
    (b) Average imaginary part.
    The averaging has been done over the time interval from $t = \SI{50}{au}$ to $t = \SI{4989}{au}$.}
    \label{fig:sal_energy}
\end{figure}

The physical energy $E = \mathrm{Re}(\mathcal{H}) = \mathrm{Re} \elm{\Psi'}{H}{\Psi}$ [Fig.~\ref{fig:sal_energy_re}] is
always conserved for the fully bivariational methods, i.e. TDMVCC, oTDMVCC, and sTDMVCC,
which is of course reassuring.
In contrast, rpTDMVCC does not conserve energy, except when the cluster expansion becomes complete. 
We take this as an unfortunate symptom of the non-bivariational nature of this method. 

The imaginary part of the Hamiltonian function, $\mathrm{Im}(\mathcal{H})$, has
no experimental significance, but it reveals some interesting and fundamental differences
between methods that are \textit{complex} bivariational (based on the complex-valued Lagrangian $\mathcal{L}$)
and methods that are \textit{real} bivariational (based on the real-valued Lagrangian $\bar{\mathcal{L}}$).
For a complex bivariational method (such as TDMVCC), the dynamics is generated by the complex-valued
Hamiltonian function $\mathcal{H}$. This entails conservation of $\mathcal{H}$ and thus of $\mathrm{Im}(\mathcal{H})$,
which is confirmed by Fig.~\ref{fig:sal_energy_im}.
The dynamics of real bivariational methods (such as oTDMVCC and sTDMVCC) is generated by
the real-valued Hamiltonian function $\mathrm{Re}(\mathcal{H})$, which implies
conservation of $\mathrm{Re}(\mathcal{H})$. We cannot expect $\mathrm{Im}(\mathcal{H})$ to be conserved, 
unless the wave function reproduces the complete-active-space solution exactly. Indeed, Fig.~\ref{fig:sal_energy_im} shows that oTDMVCC
never conserves $\mathrm{Im}(\mathcal{H})$, while sTDMVCC conserves $\mathrm{Im}(\mathcal{H})$ in the
exact limit. We remark that rpTDMVCC behaves much like sTDMVCC, but in this case the analysis is less clear
since rpTDMVCC is not fully bivariational in either the real or the complex sense.
\newpage

\newpage

\subsection{5D \textit{trans}-bithiophene}

The third and final example is the $S_1 \rightarrow S_0$ transition in
a reduced-dimensional \textit{trans}-bithiophene model that
includes five modes labelled as $Q_{10}$, $Q_{12}$, $Q_{19}$, $Q_{34}$, and $Q_{41}$
(see Ref.~\citenum{madsenVibrationallyResolvedEmission2019} for details on the selection of modes).
We obtain the initial state as the \ac{vscf} ground state on the $S_1$ surface. The wave packet
is then placed on the $S_0$ surface and propagated for a total time of \SI{20000}{au} ($\sim$\SI{484}{fs})
at the MCTDH, rpTDMVCC[2--5], oTDMVCC[2--5], sTDMVCC[2--5], and TDMVCC[2--5] levels with $N = 30$ and $N_{\!\subA} = 4$.
Once again, we limit the number of integrator steps in order to catch calculations that become unstable.

\begin{figure}[H]
    \centering

    \begin{subfigure}[c]{0.75\textwidth}
        \centering
        \caption{Average error}
        \includegraphics[width=\textwidth]{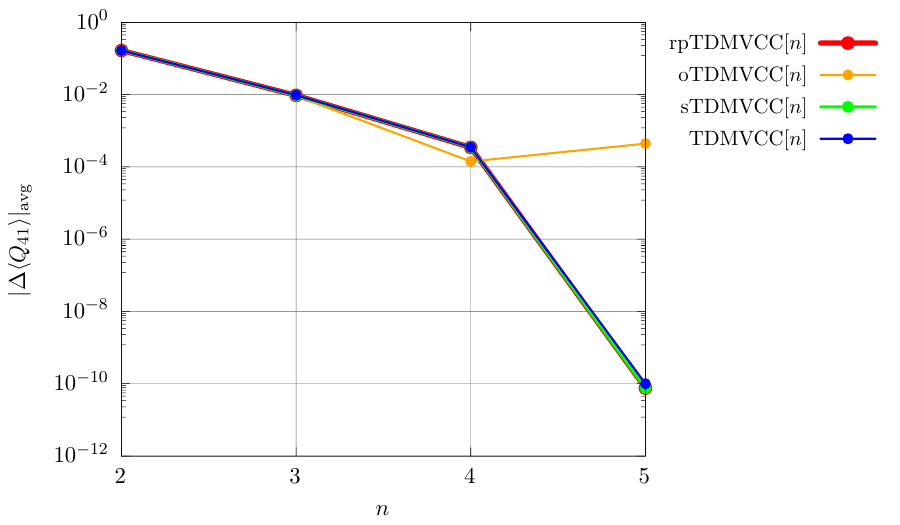}
        \label{fig:tbithio_Q41_diff}
    \end{subfigure}

    \begin{subfigure}[c]{0.75\textwidth}
        \centering
        \caption{Average imaginary vs. real part}
        \includegraphics[width=\textwidth]{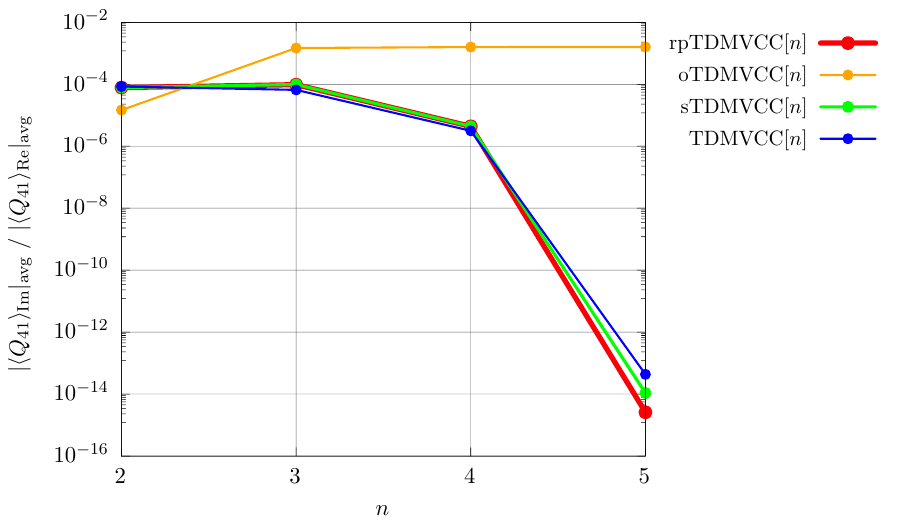}
        \label{fig:tbithio_Q41_re_im}
    \end{subfigure}

    \caption{Expectation value $\langle Q_{41} \rangle$ for 5D \textit{trans}-bithiophene 
    at the rpTDMVCC[2--5], sTDMVCC[2--5], oTDMVCC[2--5] and TDMVCC[2--5] levels.
    (a) Average error relative to MCTDH. (b) Average imaginary part divided by average real part.
    The averaging has been done over the time interval from $t = \SI{100}{au}$ to $t = \SI{12018}{au}$.}
    \label{fig:tbithio_Q41_diff_re_im}
\end{figure}

The overall behavior of 5D \textit{trans}-bithiophene 
is very similar to that of 6D salicylaldimine with respect to
stability (Fig.~S51--S54), expectation values (Fig.~S55--S64),
and energy conservation (Fig.~S65). 
As an example, we show the convergence of the expectation value of the
displacement coordinate $Q_{41}$ (inter-ring carbon--carbon stretch) 
(Fig.~\ref{fig:tbithio_Q41_diff_re_im}).
\newpage

\section{SUMMARY AND OUTLOOK} \label{sec:summary}
We have introduced a new formulation of time-dependent \ac{cc} with adaptive basis functions and division of the
one-particle space into active and secondary parts.
The formalism 
is fully bivariational and converges to the appropriate limit (i.e. MCTDH/MCTDHF) in a numerically stable manner.
The key idea of the theory is to parameterize the time-dependent basis in terms of strictly separate interspace and intraspace
transformations that change the active space itself and the active basis within that space, respectively.
Choosing a unitary interspace transformations and a non-unitary intraspace transformation
guarantees that the active bra and ket functions are biorthogonal while spanning the same space.
The former ensures convergence to the correct limit, while the latter ensures numerical stability.
The theory covers electron as well as vibrational dynamics, but our main interest lies
with the vibrational problem to which we specialize the theory. The resulting method, which is called \ac{stdmvcc},
is implemented, benchmarked, and compared to three similar \ac{cc} methods that have previously
been considered by our group. Table~\ref{tab:comparison_of_methods} summarizes our findings, 
and it is clear that only \ac{stdmvcc} succeeds in combining the
four attractive properties we have listed. In addition, we find that the split unitary/non-unitary basis
set parameterization allows a transparent and conceptually appealing understanding of adaptive basis sets in \ac{cc} theory.
Although we have not presented numerical results for the electronic structure problem,
we expect that the theory will also prove useful in this context.

We currently have
an efficient implementation of \ac{stdmvcc} that covers the two-mode coupling case ($T = T_2$ and $H = H_1 + H_2$),
and although this is likely sufficient in many cases, we believe that a
polynomial-scaling implementation that allows higher-order couplings
is required for broader applicability of the method. Such an implementation would involve
terms (such as mean fields) that do not occur in \ac{cc} ground and response theory and
are therefore not efficiently computable in our program at the present time. 
Although the techniques of Refs.~\citenum{seidlerAutomaticDerivationEvaluation2009} and \citenum{seidlerVibrationalCoupledCluster2011} can likely be used,
we expect this will be a significant undertaking.
With an efficient and general implementation at hand, we believe that the attractive combination
of a compact, evolving basis and a correlated \ac{cc} description offered by the
sTDMVCC approach will allow dynamics studies that were previously out of reach,
e.g. in the context of photochemistry. Extending the theory to also cover non-adiabatic dynamics
is an important topic for future research.

In this paper, we have exclusively considered real-time time-dependent wave functions, but it is of course
possible to employ an \ac{stdmvcc}-like \textit{Ansatz} in ground state or response theory, which are possible subjects
for future development. An \ac{stdmvcc}-type ground state may be of interest in its own right, and
it is certainly attractive as an initial state for time-dependent simulations.

\setlength{\tabcolsep}{12pt}
\begin{table}[H]
    \centering
    \caption{Overview of different variants of time-dependent vibrational coupled cluster with adaptive basis functions.
    The column `Simple' indicates whether the constraint equations involve mode-mode and up-down coupling or not.
    We currently have no formal proof that rpTDMVCC converges to MCTDH, but numerical evidence strongly suggests
    that this is in fact the case.
    }
    \begin{threeparttable}[t]
    \begin{tabular}{
    r
    c
    c
    c
    c
    }
\toprule
Method & Bivariational & Convergent & Stable & Simple \\
\midrule
  TDMVCC$^*$                 & \ding{51} &  \ding{51}  & \ding{55} & \ding{51} \\
rpTDMVCC\textsuperscript{\S} & \ding{55} & (\ding{51}) & \ding{51} & \ding{51} \\
 oTDMVCC$^\dagger$           & \ding{51} &  \ding{55}  & \ding{51} & \ding{55} \\
 sTDMVCC$^\ddagger$          & \ding{51} &  \ding{51}  & \ding{51} & \ding{51} \\
\bottomrule           
\end{tabular}
\begin{tablenotes}\footnotesize
  \item[$*$]        Refs.~\citenum{madsenTimedependentVibrationalCoupled2020} and \citenum{hojlundBivariationalTimedependentWave2022}. Analogous to \ac{oatdcc}.\cite{kvaalInitioQuantumDynamics2012}
  \item[\S]         Ref.~\citenum{hojlundBivariationalTimedependentWave2022}.
	\item[$\dagger$]  Ref.~\citenum{hojlundTimedependentCoupledCluster2024a}. Analogous to \ac{tdocc}.\cite{satoCommunicationTimedependentOptimized2018}
	\item[$\ddagger$] This work. 
\end{tablenotes}%
\end{threeparttable}%
\label{tab:comparison_of_methods}%
\end{table}%

\newpage
\section*{SUPPLEMENTARY MATERIAL} \label{sec:supplementary_material}
The supplementary material contains additional figures for water,
6D salicylaldimine, and 5D \textit{trans}-bithiophene.

\section*{ACKNOWLEDGEMENTS}
O.C. acknowledges support from the Independent Research Fund Denmark through Grant No. 1026-00122B.
This work was funded by the Danish National Research Foundation (Grant No. DNRF172) through the Center of Excellence for Chemistry of Clouds.
The numerical results presented in this paper were obtained at the Centre for Scientific Computing Aarhus (CSCAA).
The authors thank Dr. Simen Kvaal (University of Oslo) and Dr. Alberto Zoccante (Universit\`a  del Piemonte Orientale) for helpful and enlightening discussions.

\section*{AUTHOR DECLARATIONS}
\subsection*{Conflict of Interest}
The authors have no conflicts to disclose.

\subsection*{Author Contributions}
\textbf{Mads Greisen Højlund}: 
Conceptualization (lead);
Data curation (lead);
Formal analysis (lead);
Investigation (lead);
Software (lead);
Visualization (lead);
Writing -- original draft (lead);
Writing -- review \& editing (equal).
%
\textbf{Ove Christiansen}:
Conceptualization (supporting);
Formal analysis (supporting);
Funding acquisition (lead);
Project administration (lead);
Supervision (lead);
Writing -- review \& editing (equal).

\section*{DATA AVAILABILITY}
The data that supports the findings of this study are available within the article and its supplementary material.

\appendix

\section{PARTIALLY TRANSFORMED MEAN FIELDS} \label{appendix:half_transformed_mean_fields}
\setcounter{equation}{0}
\renewcommand{\theequation}{\ref{appendix:half_transformed_mean_fields}\arabic{equation}}
We start by considering the generalized Fock matrix of Eq.~\eqref{eq:F_tilde_prime_elements}:
\begin{align} \label{eq:Ftildeprime_rewrite_A}
    \Ftildeprime{m}{q}{p} 
    &= \elm{\Psi'}{[H, \creatilde{m}{p}] \annitilde{m}{q}}{\Psi} \nn
    &= \elm{\psi'}{e^{-\hat{\kappa}} [H, \creatilde{m}{p}] \annitilde{m}{q} e^{\hat{\kappa}} }{\psi} \nn
    &= {\sum_{\bar{p} \bar{q}}}^{(m)} \exp(\mathbf{K}^m)_{q \bar{q}}  \, \elm{\psi'}{[\bar{H}, \creatilde{m}{\bar{p}}] \annitilde{m}{\bar{q}}}{\psi} \exp(-\mathbf{K}^m)_{\bar{p} p} \nn
    &= {\sum_{\bar{p} \bar{q}}}^{(m)} \exp(\mathbf{K}^m)_{q \bar{q}}  \, \elm{\psi'}{ (e^{-\hat{\kappa}} e^{\hat{\kappa}}) [\bar{H}, \creatilde{m}{\bar{p}}] (e^{-\hat{\kappa}} e^{\hat{\kappa}}) \annitilde{m}{\bar{q}}  (e^{-\hat{\kappa}} e^{\hat{\kappa}}) }{\psi} \exp(-\mathbf{K}^m)_{\bar{p} p}\nn
    &= {\sum_{\bar{p} \bar{q}}}^{(m)} \exp(\mathbf{K}^m)_{q \bar{q}}  \, \elm{\breve{\psi}'}{ [H, \creabreve{m}{\bar{p}}] \bbreve{m}{\bar{q}}   }{\breve{\psi}} \exp(-\mathbf{K}^m)_{\bar{p} p}
\end{align}
At the third equality, we have performed the similarity transform, which
has the effect of transforming the Hamiltonian integrals [see Eq.~\eqref{eq:H_bar_def}]
as well as the creators and annihilators. Note that we have not absorbed the exponentials
into the basic operators, so everything is still expressed with respect to the tilde basis at this point.
At the fourth equality, we simply insert identities $1 = e^{-\hat{\kappa}} e^{\hat{\kappa}}$.
At the fifth equality, we absorb the exponentials into the creators and annihilators, thus expressing
everything with respect to the doubly transformed basis (the breve basis). We have also used 
\begin{align}
    e^{\hat{\kappa}} \bar{H} e^{-\hat{\kappa}}
    = H
    = \sum_{\mathbf{m}} {\sum_{\mathbf{p} \mathbf{q}}}^{(\mathbf{m})} \breve{H}^{\mathbf{m}}_{\mathbf{p} \mathbf{q}} \Ebreve{\mathbf{m}}{\mathbf{p}}{\mathbf{q}}.
\end{align}
Comparing the third and fifth lines in Eq.~\eqref{eq:Ftildeprime_rewrite_A}, 
we have simply relabelled the basis without changing the Hamiltonian integrals
or the amplitudes, which obviously leaves the matrix element unchanged. 
We now substitute the doubly transformed creator,
\begin{align}
    \creabreve{m}{\bar{p}} = {\sum_{\alpha}}^{(m)} \crea{m}{\alpha} \big[\mathbf{V}^m \exp(\mathbf{K}^m)\big]_{\alpha \bar{p}},
\end{align}
and simplify:
\begin{align}
    \Ftildeprime{m}{q}{p} 
    &= {\sum_{\alpha \bar{q}}}^{(m)} \exp(\mathbf{K}^m)_{q \bar{q}}  \, \elm{\breve{\psi}'}{ [H, \crea{m}{\alpha}] \bbreve{m}{\bar{q}}   }{\breve{\psi}} \vplain{m}{\alpha}{p} \nn
    &\equiv {\sum_{\alpha \bar{q}}}^{(m)} \exp(\mathbf{K}^m)_{q \bar{q}}  \, \Fbreveprime{m}{\bar{q}}{\alpha} \vplain{m}{\alpha}{p}.
\end{align}
Similarly,
\begin{align}
    \Ftildeplain{m}{q}{p}
    &= \elm{\Psi'}{\creatilde{m}{p} [\annitilde{m}{q}, H]}{\Psi} \nn
    &= \sum_{\alpha \bar{p}} \vplainconj{m}{\alpha}{q} \elm{\breve{\psi}'}{\creabreve{m}{\bar{p}} [\crea{m}{\alpha}, H]}{\breve{\psi}} \exp(-\mathbf{K}^m)_{\bar{p}p} \nn
    &\equiv \sum_{\alpha \bar{p}} \vplainconj{m}{\alpha}{q} \Fbreveplain{m}{\alpha}{\bar{p}} \exp(-\mathbf{K}^m)_{\bar{p}p}
\end{align}
The final expressions perhaps appear somewhat unintuitive
at first glance, but their structure can in fact be explained.
As an example, the element $\Ftildeplain{m}{q}{p}$ has two indices in the tilde basis,
whereas the element $\Fbreveplain{m}{\alpha}{\bar{p}}$ has one
index in the primitive basis and one index in the doubly transformed breve basis.
To go from the primitive basis to the tilde basis, one transforms by $\mathbf{V}^{m\dagger}$,
and to go from the breve basis to the tilde basis, one counter transforms by $\exp(-\mathbf{K}^m)$.
The final expressions have the additional benefit are computable in our existing code
as they stand.
Dropping the mode index from now on, we have 
\begin{subequations}
\begin{align}
    \tilde{\mathbf{F}}^{\prime} &= \exp(\mathbf{K})      \breve{\mathbf{F}}^{\prime}  \mathbf{V}, \\
    \tilde{\mathbf{F}}          &= \mathbf{V}^{\dagger}  \breve{\mathbf{F}}           \exp(-\mathbf{K}),
\end{align}
\end{subequations}
so according to Eqs.~\eqref{eq:F_check_elements},
\begin{subequations} \label{eq:Fcheck_from_Fbreve}
\begin{align}
    \check{\mathbf{F}}^{\prime} &= \exp(\mathbf{K})   \breve{\mathbf{F}}^{\prime}, \\
    \check{\mathbf{F}}          &= \breve{\mathbf{F}} \exp(-\mathbf{K}).
\end{align}
\end{subequations}
It is useful to separate the Hamiltonian into one-mode and many-mode parts, i.e.
$H = H_1 + H_{\subR}$. This induces a similar separation of $\breve{\mathbf{F}}^{\prime}$ and $\breve{\mathbf{F}}$,
and it is not hard to show that
\begin{subequations}
\begin{align}
    \breve{\mathbf{F}}^{\prime}_1 &= \mathbf{D}            \breve{\mathbf{H}}^{\prime}_1, \\
    \breve{\mathbf{F}}_1          &= \breve{\mathbf{H}}_1  \mathbf{D},
\end{align}
\end{subequations}
where the half-transformed one-mode integrals are given by
\begin{subequations}
\begin{align}
    \breve{\mathbf{H}}^{\prime}_1  &= \exp(-\mathbf{K}) \mathbf{V}^{\dagger} \mathbf{H}_1, \\
    \breve{\mathbf{H}}_1           &= \mathbf{H}_1      \mathbf{V}           \exp(\mathbf{K}).
\end{align}
\end{subequations}
The density matrix $\mathbf{D}$ is defined with respect to the untransformed states, i.e.
\begin{align}
    D^m_{qp} = \elm{\psi'}{\Etilde{m}{p}{q}}{\psi}.
\end{align}
Equations~\eqref{eq:Fcheck_from_Fbreve} now read
\begin{subequations}
    \begin{align}
        \check{\mathbf{F}}^{\prime} 
        &= \exp(\mathbf{K})      \breve{\mathbf{F}}^{\prime} \nn
        &= \exp(\mathbf{K}) \big( \breve{\mathbf{F}}^{\prime}_{1} + \breve{\mathbf{F}}^{\prime}_{\!\subR} \big) \nn
        &= \exp(\mathbf{K}) \big( \mathbf{D}  \exp(-\mathbf{K})  \mathbf{V}^{\dagger} \, \mathbf{H}_1 + \breve{\mathbf{F}}^{\prime}_{\!\subR} \big) \nn
        &= \bm{\rho} \, \mathbf{V}^{\dagger}  \mathbf{H}_1 + \exp(\mathbf{K}) \breve{\mathbf{F}}^{\prime}_{\!\subR}, \\
        \check{\mathbf{F}}        
        &= \breve{\mathbf{F}}  \exp(-\mathbf{K}) \nn
        &= \big( \breve{\mathbf{F}}_{1} + \breve{\mathbf{F}}_{\!\subR}  \big) \exp(-\mathbf{K}) \nn
        &= \big( \mathbf{H}_1  \mathbf{V} \exp(\mathbf{K})  \mathbf{D} + \breve{\mathbf{F}}_{\!\subR}  \big) \exp(-\mathbf{K}) \nn
        &= \mathbf{H}_1  \mathbf{V} \bm{\rho} + \breve{\mathbf{F}}_{\!\subR} \exp(-\mathbf{K}).
    \end{align}
\end{subequations}
Recalling that $\mathbf{H}_1$ is Hermitian since the primitive basis is orthonormal, Eq.~\eqref{eq:VA_dot} now becomes
\begin{align} \label{eq:VA_dot_appendix}
    i \dot{\mbf{V}}_{\!\!\subA} 
    &= \tfrac{1}{2} \mbf{Q} \big( \check{\mbf{F}}_{\!\!\subA} + \check{\mbf{F}}^{\prime \dagger}_{\!\!\subA} \big) \, \mathbb{H}[\bm{\rho}_{\!\subA\subA}]^{-1}  \nn
    &=
    \begin{multlined}[t]
        \tfrac{1}{2} \mbf{Q} \big( \mathbf{H}_1 \mathbf{V}_{\!\!\subA} \, {\bm{\rho}_{\!\subA\subA}} + \mathbf{H}_1 \mathbf{V}_{\!\!\subA} \, \bm{\rho}^{\dagger}_{\!\subA\subA}  \big) \, \mathbb{H}[\bm{\rho}_{\!\subA\subA}]^{-1} \\
        +  \tfrac{1}{2} \mbf{Q} \big( \breve{\mathbf{F}}_{\!\subR,\subA} \mathbf{W}_{\!\!\subA\subA} + [ \mathbf{U}_{\!\subA\subA} \breve{\mathbf{F}}^{\prime}_{\!\subR,\subA} ]^{\dagger}  \big) \, \mathbb{H}[\bm{\rho}_{\!\subA\subA}]^{-1} 
    \end{multlined} \nn
    &= \mathbf{Q} \, \mathbf{H}_1 \mathbf{V}_{\!\!\subA} 
    +  \tfrac{1}{2} \mbf{Q} \big( \breve{\mathbf{F}}_{\!\subR,\subA} \mathbf{W}_{\!\!\subA\subA}  + [ \mathbf{U}_{\!\subA\subA}  \breve{\mathbf{F}}^{\prime}_{\!\subR,\subA} ]^{\dagger}  \big) \, \mathbb{H}[\bm{\rho}_{\!\subA\subA}]^{-1} ,
\end{align}
which is the concrete expression we implement. 

\section{THE ELECTRONIC STRUCTURE CASE: ORBITAL EQUATIONS} \label{appendix:electronic_structure}
\setcounter{equation}{0}
\renewcommand{\theequation}{\ref{appendix:electronic_structure}\arabic{equation}}
The pertinent orbital equations are formally identical to those of Sato et al.,\cite{satoCommunicationTimedependentOptimized2018}
except for the lack of a $P$-space term and the presence of the $\hat{\kappa}$ transformation in the densities.
Using $o,p,q,r,s$ for the active orbitals, the equations read
\begin{align} \label{eq:orbital_equation}
    i \ket{\dot{\tilde{\varphi}}_o} = Q \Big[
        h \ket{\tilde{\varphi}_o} + \frac{1}{2} \sum_{pqrs} \tilde{W}_{\!qs} \ket{\tilde{\varphi}_r} (\rho_{rspq} + \rho_{pqrs}^*) (\mathbb{H}[\bm{\raisedrho}]^{-1})_{po}
    \Big]
\end{align}
Here, we have defined a multiplicative/local mean-field operator in the orthogonal tilde basis,
\begin{align}
    \tilde{W}_{\!qs} (\mathbf{x}) = \int \tilde{\varphi}_q (\mathbf{y}) u(\mathbf{x}, \mathbf{y}) \tilde{\varphi}_s (\mathbf{y}) \dd{\mathbf{y}},
\end{align}
and the two-electron densities
\begin{align} \label{eq:transformed_two_electron_density}
    \rho_{rspq} 
    &= \elm{\Psi'}{ \creatildeelec{p} \creatildeelec{q} \annitildeelec{s} \annitildeelec{r} }{\Psi} \nn
    &= \elm{\psi'}{ e^{-\hat{\kappa}} \creatildeelec{p} \creatildeelec{q} \annitildeelec{s} \annitildeelec{r} e^{\hat{\kappa}} }{\psi} \nn
    &= \sum_{\bar{p} \bar{q} \bar{r} \bar{s}} 
    U_{r\bar{r}} U_{s\bar{s}}
    \elm{\psi'}{ \creatildeelec{\bar{p}} \creatildeelec{\bar{q}} \annitildeelec{\bar{s}} \annitildeelec{\bar{r}} }{\psi}
    W_{\bar{p}p} W_{\bar{q} q} \nn
    &\equiv \sum_{\bar{p} \bar{q} \bar{r} \bar{s}} 
    U_{r\bar{r}} U_{s\bar{s}} P_{\bar{r} \bar{s} \bar{p} \bar{q}} W_{\bar{p}p} W_{\bar{q} q}.
\end{align}
Equation~\eqref{eq:transformed_two_electron_density} indicates a possible implementation strategy where the $\hat{\kappa}$
transformation is absorbed into the two-electron density. A different scheme is, however, also possible.
Focusing on the last term in Eq.~\eqref{eq:orbital_equation}, this reads as
\begin{align} \label{eq:mean_field_term_absorbed_into_H}
    \frac{1}{2} \sum_{qrs} \tilde{W}_{\!qs} \ket{\tilde{\varphi}_r} (\rho_{rspq} + \rho_{pqrs}^*)
    &= \frac{1}{2} \sum_{qrs} \tilde{W}_{\!qs} \ket{\tilde{\varphi}_r} \rho_{rspq}
    +  \frac{1}{2} \sum_{qrs} \Big[ \! \bra{\tilde{\varphi}_r} \tilde{W}_{\!sq} \, \rho_{pqrs} \Big]^* \nn
    &= 
    \begin{multlined}[t]
        \frac{1}{2} \sum_{qrs} \sum_{\bar{p} \bar{q} \bar{r} \bar{s}}  \tilde{W}_{\!qs} \ket{\tilde{\varphi}_r} U_{r\bar{r}} U_{s\bar{s}} P_{\bar{r} \bar{s} \bar{p} \bar{q}} W_{\bar{p}p} W_{\bar{q} q} \\
     +  \frac{1}{2} \sum_{qrs} \sum_{\bar{p} \bar{q} \bar{r} \bar{s}}  \Big[ \! \bra{\tilde{\varphi}_r} \tilde{W}_{\!sq} \, U_{p\bar{p}} U_{q\bar{q}} P_{\bar{p} \bar{q} \bar{r} \bar{s}} W_{\bar{r}r} W_{\bar{s} s} \Big]^*
    \end{multlined} \nn
    &= 
    \begin{multlined}[t]
        \frac{1}{2}  \sum_{\bar{p} \bar{q} \bar{r} \bar{s}}  \breve{W}_{\!\bar{q} \bar{s}} \ket{\breve{\varphi}_{\bar{r}}} P_{\bar{r} \bar{s} \bar{p} \bar{q}} W_{\bar{p}p} \\
     +  \frac{1}{2}  \sum_{\bar{p} \bar{q} \bar{r} \bar{s}}  \Big[ \! \bra{\breve{\varphi}_r} \breve{W}_{\!sq} \, U_{p\bar{p}} P_{\bar{p} \bar{q} \bar{r} \bar{s}} \Big]^*
    \end{multlined}
\end{align}
with
\begin{align}
    \breve{W}_{\!qs} (\mathbf{x}) = \int \breve{\varphi}'_q (\mathbf{y}) u(\mathbf{x}, \mathbf{y}) \breve{\varphi}_s (\mathbf{y}) \dd{\mathbf{y}}.
\end{align}
Equation~\eqref{eq:mean_field_term_absorbed_into_H} appears like a symmetrization of the two-electron $Q$-space terms from the \ac{oatdcc}\cite{kvaalInitioQuantumDynamics2012} orbital equations,
apart from the presence of an extra one-index transformation in each term. 
This is completely analogous to the situation in Appendix~\ref{appendix:half_transformed_mean_fields}.

\newpage

\end{document}